\documentclass[pre, twocolumn, preprintnumbers, amsmath, amssymb, nofootinbib, floatfix, superscriptaddress]{revtex4}

\usepackage{graphicx, bm, xcolor, bbold}
\usepackage{enumerate}
\usepackage[normalem]{ulem} 
\usepackage[breaklinks]{hyperref}

\makeatletter
\def\graphicscale{\twocolumn@sw{0.3}{0.4}}
\def\graphicthreescale{\twocolumn@sw{0.3}{0.4}}

\begin{document}

\title{Kibble-Zurek dynamics across the first-order quantum
  transitions\\ of quantum Ising chains in the thermodynamic limit}

\author{Andrea Pelissetto}
\altaffiliation{Authors are listed in alphabetic order.}
\affiliation{Dipartimento di Fisica dell'Universit\`a di Roma
  ``La Sapienza" and INFN, Sezione di Roma I, I-00185 Roma, Italy}

\author{Davide Rossini}
\altaffiliation{Authors are listed in alphabetic order.} 
\affiliation{Dipartimento di Fisica dell'Universit\`a di Pisa
  and INFN, Largo Pontecorvo 3, I-56127 Pisa, Italy}

\author{Ettore Vicari}
\altaffiliation{Authors are listed in alphabetic order.}
\affiliation{Dipartimento di Fisica dell'Universit\`a di Pisa,
  Largo Pontecorvo 3, I-56127 Pisa, Italy}

\date{\today}

\begin{abstract}
  We study the out-of-equilibrium Kibble-Zurek (KZ) dynamics in
  quantum Ising chains in a transverse field, driven by a
  time-dependent longitudinal field $h(t)=t/t_s$ ($t_s$ is the time
  scale of the protocol), across their first-order quantum transitions
  (FOQTs) at $h=0$.  The KZ protocol starts at time $t_i<0$ from the
  negatively magnetized ground state for $h_i = t_i/t_s<0$. Then, the
  system evolves unitarily up to a time $t_f > 0$, such that the
  magnetization of the state at time $t_f$ is positive.  In
  finite-size systems, the KZ dynamics develops out-of-equilibrium
  finite-size scaling (OFSS) behaviors. Their scaling variables depend
  either exponentially or with a power law on the size, depending on
  the boundary conditions (BC). The OFSS functions can be computed in
  effective models restricted to appropriate low-energy (magnetized
  and/or kink) states.  The KZ scaling behavior drastically changes in
  the thermodynamic limit (TL), defined as the infinite-size limit
  keeping $t$ and $t_s$ fixed, which appears substantially unrelated
  with the OFSS regime, because it involves higher-energy multi-kink
  states, which are irrelevant in the OFSS limit.  The numerical analyses
  of the KZ dynamics in the TL show the emergence of a quantum
  spinodal-like scaling behavior at the FOQTs for all considered BC, which
  is independent of the BC. The longitudinal magnetization changes sign
  at $h(t)=h_\star>0$, where $h_\star$ decreases with increasing
  $t_s$, as $h_\star\sim 1/\ln t_s$.  Moreover, in the large-$t_s$
  limit, the time-dependence of the magnetization is described by a
  universal function of $\Omega = t/\tau_s$, with $\tau_s = t_s/\ln
  t_s$.
  \end{abstract}

\maketitle

\section{Introduction}
\label{intro}

In many-body systems, the time variation of one of the parameters
across phase transitions generally gives rise to an out-of-equilibrium
dynamics, even in the limit in which time changes are infinitely slow.
These phenomena can be observed at classical transitions driven by
thermal fluctuations and at zero-temperature quantum transitions
driven by quantum fluctuations, both at continuous and first-order
transitions.

Critical out-of-equilibrium behaviors arise in Kibble-Zurek (KZ)
protocols~\cite{Kibble-80,Zurek-96}, in which a system parameter $r$
(for example, the reduced temperature at thermal phase transitions)
varies linearly as $r(t)=t/t_s$ across the transition point $r=r_c=0$,
where $t_s$ is a time scale.  Out-of-equilibrium scaling behaviors
emerge in the large-$t_s$ limit, with critical exponents related with
the length-scale critical exponent $\nu$ and the dynamic exponent $z$
that characterizes the critical slowing down of the long-distance
modes at the transition.  The first studies focused on many-body
systems in the (infinite-volume) thermodynamic limit (TL)---see, e.g.,
Refs.~\cite{Kibble-76,Kibble-80,Zurek-85,Zurek-96,ZDZ-05,PG-08,
  PSSV-11,CEGS-12}. Then, the theoretical analyses were extended to
finite systems, which show a finite-size scaling (FSS) behavior---see,
e.g., Refs.~\cite{RV-21,PRV-18-def,RV-20,TV-22,DV-23}. In finite
systems, the interplay between the out-of-equilibrium dynamic features
and the size $L$ of the system gives rise to an out-of-equilibrium FSS
(OFSS) behavior. At continuous transitions, the scaling laws
characterizing the KZ scaling behavior in the TL can be
straightforwardly obtained by taking the TL in the OFSS
relations~\cite{RV-21}.

The out-of-equilibrium behavior in KZ and more general quenching
protocols has also been studied at first-order classical and quantum
transitions~\cite{Binder-87,Pfleiderer-05,PV-24}---see, e.g.,
Refs.~\cite{RV-21,PV-24,MM-00,LFGC-09,AC-09,YKS-10,JLSZ-10,NIW-11,
  TB-12,ICA-14,PV-15,PV-16,PV-17,PV-17-dyn,LZ-17,PPV-18,SW-18,PRV-18,
  PRV-18-def,Fontana-19,LZW-19,PRV-20,DRV-20,CCP-21,SCD-21,
  CCEMP-22,TS-23,Surace-etal-24,PRV-25}. At first-order transitions,
the KZ dynamics appears more complex, showing diverse, and apparently
unrelated, behaviors in finite systems and in the TL.  The KZ dynamics
shows OFSS behaviors in finite systems, which depend on the boundary
conditions (BC),~\cite{RV-21,PV-24,RV-20,PRV-18-def,PV-17}, obtained
by generalizing the static FSS relations---see, e.g.,
Refs.~\cite{NN-75,FB-82,PF-83,FP-85,CLB-86,BK-90,LK-91,BK-92,VRSB-93,
  CNPV-14,CNPV-15,CPV-15,CPV-15-iswb,PRV-18-fowb,YCDS-18,RV-18}.  As
pointed out in Refs.~\cite{PV-17,PRV-25}, classical and quantum
scaling behaviors arise also in the TL, which, however, are not
related with the OFSS behaviors.

In this paper, we focus on the out-of-equilibrium KZ dynamics of
one-dimensional quantum Ising models in a transverse field $g$, driven
by a time-dependent longitudinal field $h(t)$ across their first-order
quantum transitions (FOQTs), occurring along the $h=0$ line for
sufficiently small values of $|g|$. We consider a KZ protocol in which
the field $h(t)$ varies as $h(t)=t/t_s$, where $t_s$ is the time scale
of the protocol.  Starting at $t=t_i<0$ from the ground state at
$h=h_i= h(t_i)<0$, where the longitudinal magnetization $m$ is
negative, the system evolves unitarily up to positive values of
$h(t)$, where $m(t)$ becomes eventually positive.

The out-of-equilibrium KZ dynamics in finite-size Ising chains is
known to obey OFSS laws, which crucially depend on the
BC~\cite{PRV-20,RV-21,PV-24,PRV-25}. OFSS is observed when the system
goes through one of the avoided level crossings of the lowest-energy
states, which occur close to the transition (in particular, at $h=0$
for BC preserving the ${\mathbb Z}_2$ symmetry), and when $t_s\sim
T(L)$, where $T(L)$ is the time scale of the transition occurring at
the crossing from a negatively magnetized state for $h<0$ to some
positively magnetized state for $h>0$~\cite{PRV-25}. In particular, if
the BC preserve the ${\mathbb Z}_2$ symmetry, then $T(L) \sim
L/\Delta(L)^2$, where $\Delta(L)$ is the energy difference (gap) of
the lowest states at the transition point $h=0$.  Since the size
dependence of the gap $\Delta(L)$ varies with the
BC~\cite{Pfeuty-70,PF-83,CJ-87,CNPV-14,RV-21,PV-24} (for example,
$\Delta(L) \sim e^{-bL}$ for open and periodic BC, while
$\Delta(L)\sim L^{-2}$ for antiperiodic BC), the main features of the
FSS and OFSS at FOQTs strongly depend on the
BC~\cite{RV-21,PV-24,LMSS-12,CNPV-14,CNPV-15, PRV-18-fowb,PRV-20}, at
variance with what happens for the FSS and OFSS at continuous quantum
transitions. We also mention that, as shown in Ref.~\cite{PRV-25} in
systems with periodic BC, OFSS can also be observed close to several
other values $h_k\sim 1/L$ of the field, corresponding to additional
avoided level crossings between the ${\rm wrongly}$ magnetized state
and kink-antikink states~\cite{SCD-21,PRV-25}.

While the OFSS behavior of the KZ dynamics in finite-size systems can
be considered as substantially
understood~\cite{RV-21,PV-24,PRV-20,PRV-25}, a thorough understanding
of the KZ dynamics in the TL has not been achieved yet, calling for
further investigations.  A first exploratory analysis of the KZ
dynamics of quantum Ising chains across their FOQTs in the TL was
recently reported in Ref.~\cite{PRV-25}, focusing on chains with
periodic BC.  It was argued that the KZ dynamics in the TL is actually
controlled by higher-energy multi-kink states, which are irrelevant in
the OFSS limit.  This would imply that the scaling relations may
substantially differ in the OFSS and TL regimes, unlike what happens
at the continuous quantum transition of the quantum Ising chain---see,
e.g., Ref.~\cite{RV-21}.  On the basis of the numerical
results~\cite{PRV-25}, the KZ dynamics across the magnetic FOQTs was
conjectured to develop an unrelated out-of-equilibrium logarithmic
scaling behavior in the TL.

In this paper we significantly extend the analysis of
Ref.~\cite{PRV-25}. We provide additional numerical evidence for the
existence of a quantum spinodal-like scaling behavior in the TL of the
KZ dynamics of quantum Ising chains across their FOQTs.  In
particular, we consider several different BC, with the purpose of
verifying whether the apparent scaling behavior in the infinite-size
limit of systems with periodic BC~\cite{PRV-25} is a general feature
of the KZ dynamics in the TL.  Our analyses confirm the emergence of
the spinodal-like out-of-equilibrium scaling behavior in the TL for
all the considered BC, allowing us to obtain an accurate
phenomenological characterization.  The negatively magnetized state
turns out to jump to states with positive magnetization at
$h(t)=h_\star(t_s)>0$, where $h_\star(t_s)$ approaches $h=0^+$ with
increasing $t_s$, apparently as $h_\star\sim 1/\ln t_s$. Moreover, in
the TL the time evolution of the longitudinal magnetization shows a
universal scaling behavior in terms of the scaling variable $\Omega =
t/\tau_s$ with $\tau_s = t_s/\ln t_s$. The scaling behavior turns out
to be independent of the BC.

The paper is organized as follows.  In Sec.~\ref{model} we define the
one-dimensional quantum Ising model and the KZ protocol.  In
Sec.~\ref{ofss} we outline the main features of the OFSS behavior
observed when the Ising chain is driven across a FOQT by a
time-varying longitudinal field (KZ dynamics).  Sec.~\ref{numres}
reports the numerical results for different BC, which allow us to
understand the scaling behavior of the KZ dynamics in the TL.
Finally, in Sec.~\ref{conclu} we summarize and draw our
conclusions. In App.~\ref{AppOBC} we report some detailed analysis of
the low-energy spectrum for Ising chains with open BC and opposite
fixed BC. In App.~\ref{AppOFSS} we discuss the KZ dynamics in the OFSS
limit for quantum Ising chains with opposite fixed BC.

\section{Model and dynamic protocol}
\label{model}

\subsection{The quantum Ising chain}
\label{quisch}

The nearest-neighbor quantum Ising chain in a transverse field is a
paradigmatic model showing continuous and first-order quantum
transitions. The Hamiltonian of a chain of size $L$ reads
\begin{equation}
  \hat H = - J \, \sum_{\langle x,y\rangle} \hat\sigma^{(1)}_x
  \hat\sigma^{(1)}_{y}
  - g\, \sum_x \hat\sigma^{(3)}_x - h \,\sum_x
  \hat\sigma^{(1)}_x,
  \label{hedef}
\end{equation}
where $\hat\sigma^{(\alpha)}$ are the spin-$1/2$ Pauli matrices
($\alpha=1,2,3$), the first sum is over all nearest-neighbor bonds
$\langle x,y\rangle$, while the second and the third sums are over the
$L$ sites of the chain ($x$ runs from 1 to $L$).  The Hamiltonian
parameters $g$ and $h$ represent homogeneous transverse and
longitudinal fields, respectively.  Without loss of generality, we
assume $J = 1$ and $g>0$. We also set the Planck constant $\hslash =
1$.

In the zero-temperature limit and for $g=1$, $h=0$, the
model~\eqref{hedef} undergoes a continuous quantum transition
belonging to the two-dimensional Ising universality class, separating
a disordered phase ($g>1$) from an ordered ($g<1$) one---see, e.g.,
Refs.~\cite{Sachdev-book,RV-21,CPV-14,PV-02} for more details.  For
any $g<1$, the longitudinal field $h$ drives FOQTs along the $h=0$
line, leading to a discontinuity of the (average) ground-state
longitudinal magnetization
\begin{equation}
  m = {1\over L}\sum_{x=1}^L m_x,\quad m_{x} = \langle \Psi_0(g,h) |
  \hat\sigma_{x}^{(1)}| \Psi_0(g,h)\rangle,
  \label{mxdef}
\end{equation}
where $|\Psi_0(g,h)\rangle$ is the ground state for the Hamiltonian
parameters $g$ and $h$. Indeed, the FOQT separates two different
phases characterized by opposite nonzero values of $m$,
i.e.,~\cite{Pfeuty-70}
\begin{equation}
  \lim_{h \to 0^\pm} \lim_{L\to\infty} m = \pm \,m_0(g), \quad m_0(g)
  = (1 - g^2)^{1/8}.
  \label{disco}
\end{equation}

We discuss the approach to the TL in systems with {\it (i)} periodic
BC (PBC), for which $\hat\sigma^{(\alpha)}_x = \hat\sigma^{(\alpha)}_{x+L}$;
{\it (ii)} open BC (OBC); {\it (iii)} equal fixed BC (EFBC), where one considers
a chain with two additional boundary sites ($x=0$ and $x=L+1$) and restricts
the state space to states that are eigenstates of $\hat\sigma^{(1)}_0$
and $\hat\sigma^{(1)}_{L+1}$ with the same eigenvalue; {\it (iv)} opposite
fixed BC (OFBC), where again one considers a chain of length $L+2$,
but now states are eigenstates of $\hat\sigma^{(1)}_0$ and
$\hat\sigma^{(1)}_{L+1}$ with opposite eigenvalue; {\it (v)} antiperiodic
BC (ABC), for which $\hat\sigma^{(\alpha)}_x = - \hat\sigma^{(\alpha)}_{x+L}$.
Systems with PBC, OBC, and ABC are ${\mathbb Z}_2$ symmetric. In the OFBC case,
the system is symmetric under transformations that combine ${\mathbb Z}_2$
reflections and the space inversions with respect to the
center of the chain.  Translation invariance is preserved by PBC and
ABC, giving rise to selection rules in momentum space.

As already mentioned, different BC are known to lead to
different OFSS behaviors~\cite{PV-24, RV-21, PRV-18, PRV-20, TV-22,
  TS-23}. For example, scaling variables depend exponentially on the
size of the system for OBC, EFBC, and PBC, while a power dependence
emerges when considering OFBC and ABC. This is due to the different
nature of the lowest-energy levels, which crucially depends on the BC
at FOQTs---see, e.g., Refs.~\cite{Pfeuty-70,CJ-87,CNPV-14,CNPV-15,
  CPV-15, CPV-15-iswb,PRV-18-fowb,LMSS-12,RV-21,PV-24}).

In finite systems
with PBC or OBC, the two lowest eigenstates of $\hat{H}$ are superpositions
of the magnetized states $|+\rangle$ and $|-\rangle$, satisfying
$\langle \, \pm \, | \,\hat\sigma_{x}^{(1)} \,| \, \pm \,\rangle = \pm
\,m_0$ (sufficiently far from the boundaries in the OBC case). Their
energy gap $\Delta(L)$ vanishes exponentially with increasing
$L$~\cite{Pfeuty-70, CJ-87}, as $\Delta(L)\sim e^{-c L}$, where $c$ is
a positive constant that depends on $g$ (multiplicative powers of $L$
may also be present).  The difference of the energies of the higher
excited states and of the two lowest-energy states is finite in the
infinite-volume limit.

For OFBC and ABC, the magnetized states are no longer the relevant
low-energy states. Instead, the lowest-energy states are domain walls
(kinks), which, for small values of $g$, are simply characterized by
the presence of nearest-neighbor pairs of antiparallel spins.  Energy
eigenstates are linear combinations of kink states that behave as
one-particle states with $O(L^{-1})$ momenta. Therefore, the
low-energy spectrum is characterized by energy gaps that scale as
$\Delta(L)\sim L^{-2}$~\cite{CJ-87,CNPV-14,CPV-15}.

Finally, in the EFBC case, the global ${\mathbb Z}_2$ symmetry is
broken and only one magnetized state is allowed by the BC.
In this case, observables depend smoothly on $h$ around
$h=0$ (for sufficiently small values of $|h|$), up to a value
$h_{\rm tr}\approx c/L$, where a sharp transition to the oppositely
magnetized phase occurs~\cite{PRV-18-fowb}. For each $L$, the transition
field $h_{\rm tr}(L)$ can be identified as the value of $h$ where the
energy difference between the two lowest-energy states takes its minimum.
The gap $\Delta_{\rm min}(L) = \Delta(L,h_{\rm tr})$ decreases as
$e^{-bL}$ with the system size. Note that the infinite-volume
limit and the $h\to 0$ limit do not commute.  Indeed, the gap
$\Delta(L,h)$ at $h=0$ is finite for $L\to \infty$.

\subsection{The Kibble-Zurek protocol}
\label{protocol}
  
To investigate the out-of-equilibrium behavior that arises when
crossing a FOQT, we consider a dynamic KZ protocol in which the
longitudinal field varies across the value $h=0$, for fixed $g<1$.  The
system evolves unitarily according to the Schr\"odinger equation
\begin{equation}
  i{{\rm d} \, |\Psi(t)\rangle \over {\rm d} t} = \hat H[h(t),g] \,
  |\Psi(t)\rangle ,\qquad h(t)=t/t_s,
  \label{unitdyn}
\end{equation}
where $t_s$ is a time scale.  We consider KZ protocols that start at
time $t_i=h_i \, t_s$ with $h_i<0$ fixed, from the corresponding ground
state $|\Psi(t_i)\rangle \equiv |\Psi_0(h_i,g)\rangle$ with negative
magnetization $m(t_i)$.  Then, the system evolves up to a time
$t=t_f>0$, corresponding to $h(t_f)=h_f=t_f/t_s>0$, which is
sufficiently large to obtain states $|\Psi(t)\rangle$ with positive
longitudinal magnetization. Note that, in the $t_s\to\infty$ limit
keeping $L$ fixed, the KZ evolution is adiabatic: $|\Psi(t)\rangle$
corresponds to the ground state $|\Psi_0[h(t),g]\rangle$. Instead, if
we take the $L\to\infty$ limit at fixed $t_s$, we obtain an
out-of-equilibrium dynamics for any finite $t_s$ and also in the
$t_s\to\infty$ limit.

To monitor the evolution of the system, we compute the instantaneous
local longitudinal magnetization
\begin{equation}
  m_x(t,t_s,L) = \langle\Psi(t)| \hat\sigma_{x}^{(1)}|\Psi(t) \rangle.
  \label{mxm}
\end{equation}
In particular, we consider the rescaled central magnetization
(averaged over the two central sites, since we generally consider
chains of even size $L$) and the rescaled average magnetization,
defined as
\begin{equation}
  M_c = {m_{L/2}+m_{{L/2}+1}\over 2 \, m_0},\qquad
  M = {1 \over m_0\,L} \,\sum_{x=1}^L m_x,
  \label{MMcdef}
\end{equation}
where $m_0$ is the $g$-dependent value of the longitudinal
magnetization given in~Eq.~\eqref{disco}.  For BC preserving
translation invariance, such as PBC and ABC, $M = M_c$. For systems
with boundaries, to minimize boundary effects, we mostly consider the
central magnetization $M_c$, which is computed at the sites that are
at the farthest distance from the boundaries.

In this paper, we numerically investigate the KZ protocol outlined above.
We first find the ground state of the initial Hamiltonian in the full
Hilbert space, with exact-diagonalization methods. Then we
integrate the corresponding Schr\"odinger equation~\eqref{unitdyn}
with a fourth-order Runge-Kutta algorithm.  We choose a sufficiently
small time step $dt = 2.5 \times 10^{-3}$, to ensure convergence of
all our results up to the largest considered sizes ($L = 22$) and
times ($t \sim 10^3$).  For PBC, we exploit momentum conservation,
which allows us to work in a smaller Hilbert space and to reach larger
system sizes ($L = 26$) with approximately the same computational
resources~\cite{Sandvik-10}.

To study the dynamics in the TL, we follow a two-step procedure. ({\it i})
First, we determine the large-$L$ limit at fixed time scale $t_s$, by
increasing $L$ until the central magnetization $M_c(t,t_s,L)$ appears
to approach an $L$-independent function $M_\infty(t,t_s)$.  This
limits the values of $t_s$ we can probe, because the smallest values
of $L$ at which approximately size-independent results are observed
increase rapidly with increasing $t_s$ (our numerical computations are
limited to chains with a few tens of sites). ({\it ii}) Then, we study
the behavior of the infinite-size magnetization $M_\infty(t,t_s)$ as a
function of $t_s$, looking for a scaling behavior in terms of the
variables $t$ and $t_s$, for large values of $t_s$.

\section{Out-of-equilibrium finite-size scaling}  
\label{ofss}

In this section we review the OFSS theory that characterizes
the KZ dynamics in finite-size  Ising chains driven 
across the $h=0$ FOQT.

At the FOQT at $h=0$, the low-energy properties satisfy equilibrium
FSS laws as a function of the field $h$ and of the system size
$L$~\cite{CNPV-14,PRV-18-fowb,CNPV-15,PV-24}.  When the BC preserve
the ${\mathbb Z}_2$ symmetry, so all properties are symmetric under
$h\to -h$, the relevant scaling variable is the ratio~\cite{CNPV-14}
\begin{equation}
  \Phi = {\delta E(L,h) \over \Delta(L)} =  {2 m_0 h L \over \Delta(L)},
  \label{phidef}
\end{equation}
where $\delta E(L,h)= 2 m_0 hL$ quantifies the magnetic energy
associated with the longitudinal field $h$, and $\Delta(L)$ is the
difference of the energy of the two lowest levels at $h=0$.  The
zero-temperature FSS limit corresponds to $L \to \infty$ and $h \to
0$, keeping $\Phi$ fixed.  In this limit, the ground-state
magnetization defined in~Eq.~\eqref{mxdef} behaves as~\cite{CNPV-14}
\begin{equation}
  m(L,h) \approx m_0\,{\cal M}(\Phi),
  \label{fssmag}
  \end{equation}
where ${\cal M}(\Phi)$ is a scaling function independent of $g$ (as
long as $g<1$).  An analogous FSS behavior is found for other
observables, such as the energy gap and the ground-state
fidelity~\cite{RV-18,RV-21,PV-24}. Moreover, it can be
straightforwardly extended to allow for a nonzero
temperature~\cite{RV-21}.

For the KZ dynamics, OFSS laws can be derived by extending the
equilibrium FSS relations reported above. We first define a
time-dependent scaling variable that corresponds to $\Phi$ defined in
Eq.~\eqref{phidef}~\cite{PRV-18-def,RV-21,PV-24,PRV-25}:
\begin{equation}
  \widehat \Phi \equiv {2 m_0 h(t) L \over \Delta(L)}
  = {2 m_0 t L \over t_s\Delta(L)}.
  \label{katdef}
\end{equation}
A second scaling variable associated with time~\cite{PRV-18} is given
by $\Theta \equiv t\,\Delta(L)$. Combining $\Theta$ and
$\widehat{\Phi}$, we can define the time-independent scaling variable
\begin{equation}
  \Upsilon = {\Theta \over \widehat\Phi} = {t_s \over T(L)},\qquad
  T(L)={2m_0 L\over \Delta(L)^2},
  \label{upsilondef}
\end{equation}
which is the ratio between $t_s$ and the time scale $T(L)$ that
characterizes the crossing of the transition point $h=0$ for a system
of size $L$.  The OFSS limit corresponds to $t, t_s, L\,\to\infty$,
keeping $\widehat\Phi$ and $\Upsilon$ fixed. In this limit, the
rescaled longitudinal magnetization 
scales as~\cite{PRV-18-def,PRV-20,RV-21,PV-24}
\begin{equation}
  M(t,t_s,h_i,L) \approx {\cal M}(\Upsilon, \widehat\Phi),
  \label{mtsl}
\end{equation}
independently of $h_i$ (for fixed $h_i<0$).  The adiabatic limit
corresponds to $t,t_s\to \infty$ at fixed $L$ and $t/t_s$, thus
implying $\Upsilon\to \infty$.  In this limit, ${\cal M}(\Upsilon,
\widehat\Phi)$ converges to the equilibrium FSS function defined
in~\eqref{fssmag} with $\widehat\Phi=\Phi$.

The above OFSS relations are expected to hold for any type of BC that
preserve the ${\mathbb Z}_2$ symmetry, independently of the size
dependence of the gap $\Delta(L)$.  In the EFBC case, because of the
boundary violation of the ${\mathbb Z}_2$ symmetry, both the
equilibrium FSS and the OFSS should be modified. Indeed, scaling is
not observed for $h\approx 0$, but only close to the pseudotransition
point $h = h_{\rm tr}(L) \sim 1/L$.  This requires a redefinition of
the FSS variable $\Phi$~\cite{PRV-18-fowb}.  In the EFBC case one
should use
\begin{equation}
  \Phi_{e} = {2m_0[h-h_{\rm tr}(L)] L\over \Delta_{m}(L)},
  \label{phisdef}
\end{equation}
where $\Delta_m(L) \equiv \Delta(L,h_{\rm tr})$ is the ground-state
gap for $h=h_{\rm tr}(L)$. Analogously, to define the OFSS for the KZ
dynamics, one must replace $\widehat{\Phi}$ and $\Upsilon$ defined in
Eqs.~\eqref{katdef} and~\eqref{upsilondef} with $\widehat{\Phi}_e$ and
$\Upsilon_e$, defined as~\cite{PRV-20}
\begin{eqnarray}
& \widehat\Phi_{e} = \displaystyle {2 m_0 L t_e \over t_s \Delta_{m}(L)},
\quad  & t_e = t - h_{\rm tr}(L)\,t_s, \label{efbcphi}\\
&  \displaystyle \Upsilon_{e} = { t_s \over T_{e}(L)}, \quad
        &   T_{e}(L) = {2m_0L\over \Delta_{m}(L)^2}.
  \label{efbcredef}
\end{eqnarray}
Also in this case $T_e(L)$ is the relevant time characterizing the
passage across the avoided level crossing at $h_{\rm tr}$. As
$\Delta_{m}(L)$ is exponentially small, $T_e(L)$ increases
exponentially with the system size.

The FSS and OFSS functions depend on the BC.  In the OBC and PBC case,
in which the relevant states are the two magnetized states
$|\pm\rangle$, the FSS and OFSS functions can be computed by using a
two-level effective model~\cite{CNPV-14,PRV-20,PV-24,PRV-25}.  For
example, the OFSS function associated with the longitudinal
magnetization can be computed in terms of the scaling variables
$\widehat\Phi$ and $\Upsilon$, cf.~Eq.~\eqref{mtsl}, using the
solution of the Landau-Zener two-level
problem~\cite{Landau-32,Zener-32,VG-96}.  For OBC and PBC---and, more
generally, for all ${\mathbb Z}_2$-symmetric BC with a fully
magnetized ground state---one obtains~\cite{PRV-20,RV-21,PV-24}
\begin{equation}
  {\cal M}(\Upsilon, \widehat\Phi) = -1 + \tfrac12 \Upsilon e^{-{\pi
      \Upsilon\over 8}} \left| D_{-1+i{\Upsilon\over 4}} (e^{i{3\pi\over4}}
  \widehat{\Phi}\,\Upsilon^{1\over 2}) \right|^2,
  \label{mofss}
\end{equation}
where $D_\nu(x)$ is the parabolic cylinder function~\cite{AS-1964}.
For systems with boundaries, such as OBC, Eq.~\eqref{mofss} applies to
the local longitudinal magnetization far from the boundaries, for
example to $M_c(t)$ at the central sites. Note that the two-level
approximation works for any avoided-level crossing that involves only
two states, and therefore also in the absence of ${\mathbb Z}_2$
symmetry. For EFBC, the appropriate scaling function of the
magnetization is reported in Ref.~\cite{PRV-20}.

It is important to stress that the static FSS and the OFSS only apply
in a small interval of values of $h$ of width $\Delta(L)/L$, which
shrinks to zero as $L\to \infty$. In this interval only the lowest
energy states of the spectrum for $h=0$ (the magnetized states for
PBC, OBC, EFBC, the kink states for OFBC and ABC as discussed below)
play a role.  Moreover, if $t_s \ll T(L)$, where $T(L)=L/\Delta(L)^2$
is the typical crossing time scale, the OFSS is trivial: after
crossing the transition, the system is in the same state as before
the transition.  As already noted in Ref.~\cite{PRV-25}, based on the
analysis of systems with PBC, the OFSS regime is irrelevant for the
understanding of the scaling behavior of the system in the TL for
finite values of the magnetic field. Indeed, for finite values of
$h(t)$ and large values of $L$, the longitudinal magnetic Hamiltonian
term is the relevant one, so the relevant states are completely
different from those playing a role in the OFSS regime (for instance,
one should consider all positively-magnetized multikink states for $h
> 0$).  As we shall see, the numerical data confirm this picture, also
indicating that the passage from states with $M<0$ to states with
$M>0$ occurs at $h(t)=h_\star>0$, where $h_\star\sim 1/\ln t_s$.

\section{Numerical results}
\label{numres}

We now present our numerical results for the KZ dynamics in Ising
chains for different BC. We mostly focus on the behavior in the TL, as
defined in Sec.~\ref{protocol}. Some results for Ising chains with PBC
were already reported in Ref.~\cite{PRV-25} (see its Sec. VII), up to
sizes $L=20$; in the following we also report some further results for
PBC for larger chains, up to $L=26$.  We also compare the obtained
behaviors for the various BC considered in some detail.

\subsection{Open boundary conditions}
\label{obcsec}

Along the FOQT line ($g < 1$ and $h=0$) the two lowest-energy states
for chains with OBC are combinations of the two magnetized states
$|\,\pm\,\rangle$, with an energy gap that vanishes as $L\to \infty$.
Above them, one can identify a family of single-kink states ($2L-2$
states for a chain of length $L$). The energy difference between kink
states and the lowest-energy magnetized states is finite for large
$L$, while the energy difference between two different single-kink
states is of the order of $1/L^2$ for large $L$, so they become degenerate
in the TL.  For $g\to 0$, single-kink states correspond to
\begin{equation}
  |{\rm k}_{x}\rangle = | \cdots \uparrow_{x-1} \: \uparrow_x \:
  \downarrow_{x+1} \: \downarrow_{x+2} \cdots \rangle,
  \label{kstates}
\end{equation}
with a single pair of antiparallel spins (those at sites $x$ and
$x+1$).  For positive values of $h$, the magnetized state $|+\rangle$
is the ground state of the system,  while the magnetized state
$|-\rangle$ and the single-kink states give rise to a sequence of
avoided level crossings. The spectrum is analogous to that of chains
with PBC, discussed in Ref.~\cite{PRV-25}. The only difference is the
nature of the lowest-lying states. While with OBC single-kink states
are the lowest-energy relevant excitations, with PBC
one should consider kink-antikink states, such as
\begin{equation}
  |{\rm k}_x \, \bar{\rm k}_{x+w}\rangle = | \cdots \uparrow_{x-1} \:
  \uparrow_x \: \downarrow_{x+1} \cdots \downarrow_{x+w} \:
  \uparrow_{x+w+1} \cdots \rangle.
  \label{kakstates}
\end{equation}
This analogy allows one to extend the spectrum results reported in
Ref.~\cite{PRV-25} to systems with OBC (see App.~\ref{AppOBC}).  In
particular, it follows that the low-energy spectrum of finite-size
systems with OBC shows a sequence of avoided level
crossings~\cite{SCD-21,PRV-25} between the ${\rm wrongly}$ magnetized
state and a discrete series of single-kink states, labeled by $k=1,2,
\ldots$. These are located at
\begin{equation}
  h_k(L) = {a\over L} + {b_k\over L^{5/3}} + O(L^{-2}),
  \label{hkbeh}
\end{equation}  
 where $a$
is independent of $k$ and $b_k$ increases with $k$.

Quantum Ising chains with OBC develop the OFSS outlined in
Sec~\ref{ofss} when $t_s\sim T(L)=L/\Delta(L)^{2}$, cf.
Eq.~(\ref{upsilondef}).  Since~\cite{Pfeuty-70,CJ-87}
\begin{equation}
  \Delta(L) = 2 (1-g^2) g^L [ 1 + O(g^{2L})],
  \label{deltaobc}
\end{equation}
$T(L)$ increases exponentially with the system size.  Moreover,
due to the fact that the energy difference with the higher states
is finite in the large-$L$ limit, the OFSS functions can be evaluated
in an effective two-level model~\cite{CNPV-14}: the scaling function
of the central magnetization is reported in Eq.~(\ref{mofss}).
If, on the other hand, $t_s\ll T(L)$, the system moves across the transition
point $h=0$ so fast that it is unable to jump to the positively
magnetized state. Thus, for $h >0$ the system is still in the
negatively magnetized state. This is consistent with
Eq.~(\ref{mofss}), that predicts ${\cal M}(\Upsilon,\hat{\Phi}) \to
-1$ for $\Upsilon = t_s/T(L) \to 0$.

As already discussed for PBC~\cite{PRV-25}, the presence of further
avoided crossings between single-kink states and the negatively
magnetized state gives rise to additional OFSS regimes. We can
associate a time scale $T_k(L)$ to the crossing occurring at $h_k(L)$
and observe a nontrivial OFSS behavior when $t_s$ is of the order of
$T_k(L)$. More precisely, if we assume $T(L) \gg T_1(L) \gg T_2(L) \gg
\ldots$ for sufficiently large sizes, analogously to the PBC case, the
time scale $t_s$ can be tuned in such a way to select at which avoided
crossing the system magnetization changes sign.  If $t_s \approx
T_k(L)$ and $t_s \ll T_{k-1}(L)$, the negatively magnetized state
effectively survives across the $h=0$ and the first $k-1$ avoided
level crossings, up to the one satisfying $t_s\approx T_k(L)$. When
$h(t) \approx h_k(L)$, the system jumps to a kink state with positive
magnetization. This behavior has been verified numerically in
Ref.~\cite{PRV-25} for Ising chains with PBC.  Here we do not present
results for OBC in these further OFSS regimes.

The intermediate scaling regimes occurring at the sequence of avoided
level crossings is not relevant to describe the KZ dynamics in the TL.
This is not only due to the fact that the sequence of avoided
crossings collapses towards $h=0$ in the large-$L$ limit, but it is
essentially related to the fact that the energy states relevant for
the TL are expected to be different. While only single-kink states are
relevant in the OFSS regimes, in the TL positively magnetized
multi-kink states are expected to play an important role.  Therefore,
in the TL, the KZ dynamics may show an unrelated scaling behavior, as
already put forward for PBC.

\begin{figure}[!t]
  \includegraphics[width=0.47\textwidth]{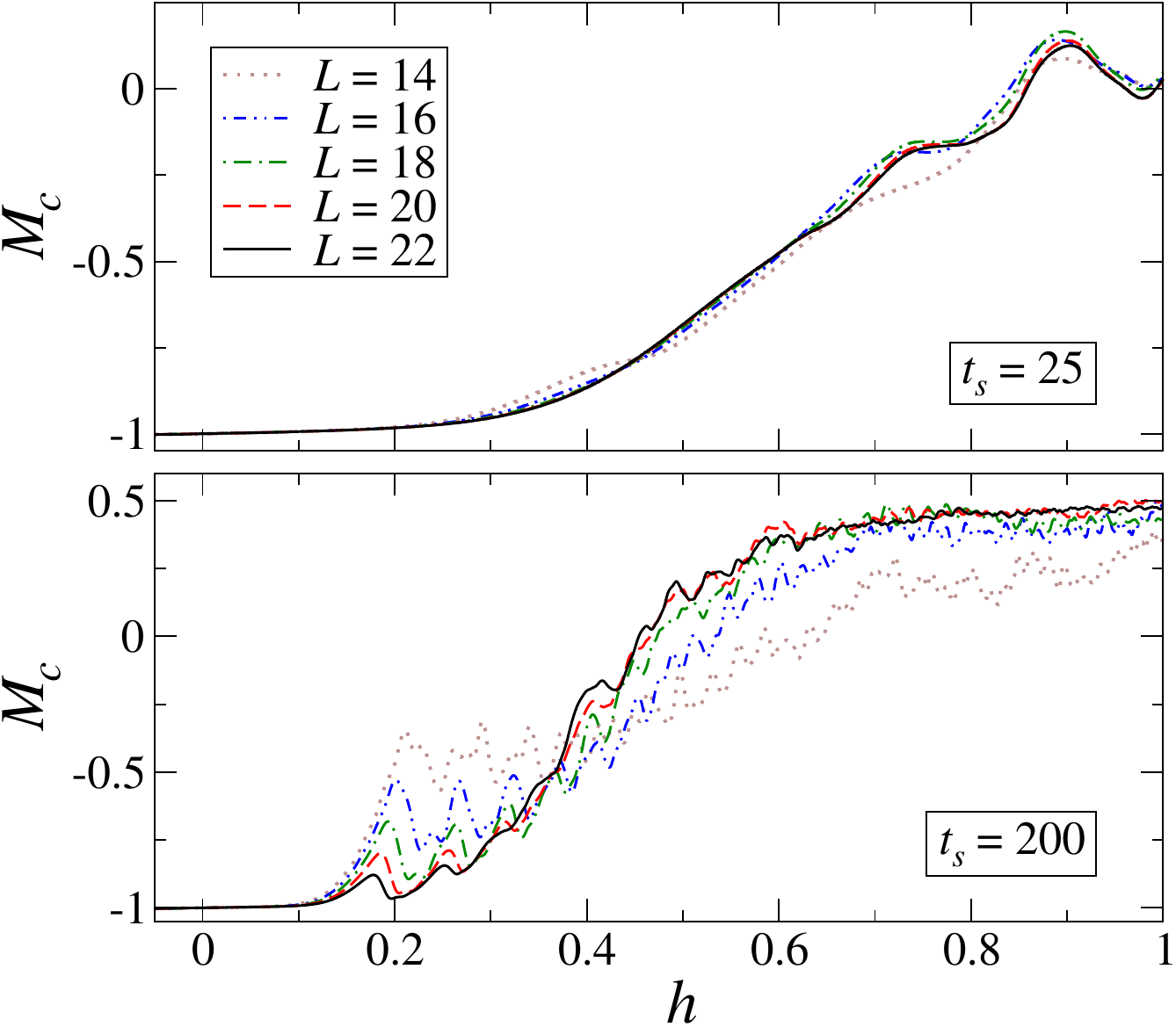}
  \caption{Rescaled central longitudinal magnetization $M_c$
    defined in~Eq.~\eqref{MMcdef}, for chains with OBC, plotted
    vs $h(t)$, for $t_s = 25$ (top) and $200$ (bottom). We report
    curves for different lattice sizes up to $L=22$ (see legend).
    With increasing $L$, the data appear to converge to an asymptotic
    curve, which we consider as the infinite-size limit at fixed $h(t)$.
    Unless otherwise specified, all numerical data shown here and in the
    following figures have been obtained fixing $g=0.5$.}
  \label{OBCinfL}
\end{figure}

To study the KZ dynamics in the TL, we analyze data for Ising chains
at fixed transverse field $g=0.5$, following the two-step procedure
outlined at the end of Sec.~\ref{protocol}.

Since systems with OBC
have boundaries, to minimize size effects we focus on the behavior of
the central longitudinal magnetization $M_c(t)$ defined in
Eq.~(\ref{MMcdef}). At first, we determine $M_c(t)$ for $L\to \infty$
at fixed $t_s$. Results are shown in Fig.~\ref{OBCinfL}, where we
display $M_c(t)$ for $t_s=25$ and $t_s=200$, and several system sizes
up to $L=22$.  We observe that the different data sets apparently
converge to an asymptotic large-$L$ curve, which provides the
time-dependent infinite-size magnetization $M_{c,\infty}(t,t_s)$ for
the given value of $t_s$.  Convergence is faster for small time scales
$t_s$. A good convergence is observed for $t_s = 25$ (top panel),
while, for $t_s=200$ (bottom panel), fast time oscillations emerge
although with an amplitude that decreases with $L$. Nonetheless, an
asymptotic behavior can be fairly recognized for chains of length $L
\approx 22$.  We have not considered larger values of $t_s$, as the TL
asymptotic behavior would be observed only for significantly longer
chains that cannot be studied with the existing computational
resources in a reasonable amount of time.

\begin{figure}[!t]
  \includegraphics[width=0.47\textwidth]{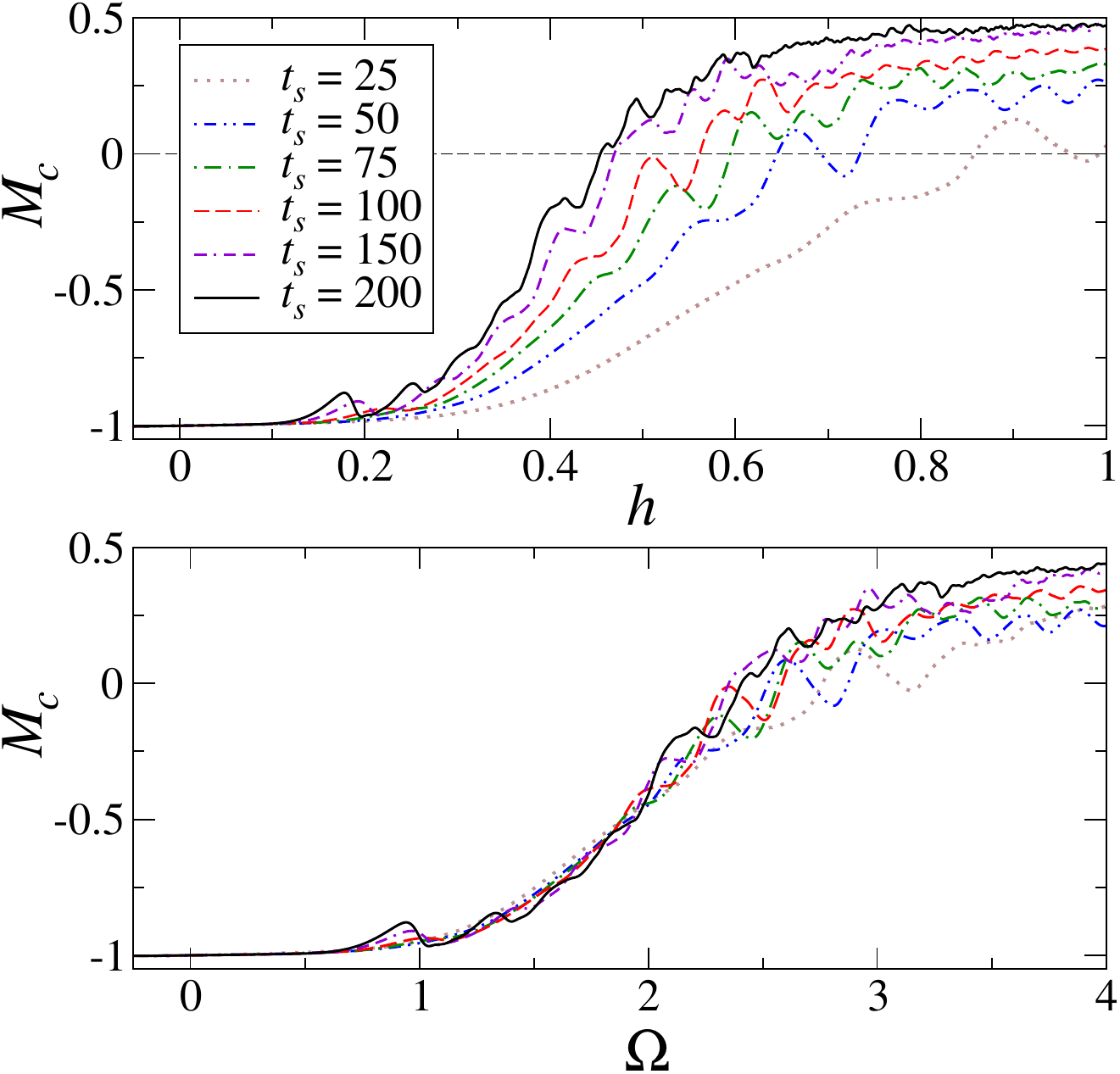}
  \caption{Rescaled central magnetization $M_c$ for chains with OBC and $L=22$,
    plotted vs $h(t)$ (top) and vs the rescaled variable
    $\Omega(t) = h(t) \, \ln t_s$ (bottom).  The curves correspond to
    different time scales, up to $t_s = 200$ (see legend). For
    these values of $t_s$ the data for chains with 22 sites provide,
    with good approximation, the time behavior of $M_c$ in the
    infinite-size limit.}
  \label{OBCtssca}
\end{figure}

In the second step, we compare the infinite-size time evolution of the
magnetization for several increasing time scales $t_s$, looking for
the emergence of a scaling behavior for large values of $t_s$
(cf.~Fig.~\ref{OBCtssca}). The top panel shows data up to $t_s = 200$,
vs $h(t)=t/t_s$. We note that $M_c$ changes sign for values of $h$
that decrease towards $h=0$ with increasing $t_s$ (see the dashed
horizontal line in the figure, which is plotted to guide the eye).  In
the bottom panel, the same data are plotted vs the scaling variable
$\Omega(t)=h(t)\,\ln t_s$, identified in Ref.~\cite{PRV-25} as the
relevant scaling variable for the KZ dynamics of large-$L$ systems
with PBC.  We observe that the curves for the infinite-size
magnetization apparently fall onto a single curve. The collapse is
evident at least up to $\Omega(t) \approx 2.5$, corresponding to
$M_c(t) \lesssim 0$.  For $\Omega \gtrsim 2.5$, time oscillations set
in, which prevent us from observing a clear data collapse. Of course,
larger time scales $t_s$ and larger (computationally inaccessible)
system sizes would be required to obtain a robust evidence of scaling
also for these larger $\Omega$ values.

These results lead us to conjecture that the central longitudinal
magnetization behaves as
\begin{eqnarray}
  &&  M_{c,\infty}(t,t_s) \approx {\cal M}_\infty(\Omega),
    \label{miscai}\\
  && \Omega(t) = {t\over \tau_s},\qquad  \tau_s={t_s\over \ln t_s},
\label{omegadef}
\end{eqnarray}
in the large-$t_s$ limit, as already conjectured and numerically
verified for systems with PBC~\cite{PRV-25}.  Note that the scaling
relation (\ref{miscai}) implies that the longitudinal magnetization
changes sign for a fixed value of $\Omega$ in the large-$t_s$ limit,
which also implies a logarithmic scaling of the corresponding
longitudinal field $h_\star$, behaving as $h_\star\sim 1/\ln t_s$.

We also note that the magnetization approaches an almost constant
value (apart from short-time fluctuations), $M_\infty(t\to\infty,t_s)
\approx 0.5$, for sufficiently large values of $\Omega$ and of $t_s$.
The significant deviation from one (i.e., from the magnetization of
the ground state for $h\to\infty$), can be explained by the large
energy excess, therefore by the fact that the KZ protocol has injected
a relatively large (of order $L$) amount of energy (the work needed to
change $h$) in the system, giving rise to a significant departure from
the ground state of the Hamiltonian at large times.

\subsection{Equal fixed boundary conditions}
\label{EFBCsec}

To implement EFBC we consider a chain with $L+2$ sites and
Hamiltonian~\eqref{hedef}, restricting the Hilbert space to states
$|s\rangle$ such that $\hat\sigma^{(1)}_0 |s\rangle = -|s\rangle$ and
$\hat\sigma^{(1)}_{L+1} |s\rangle = -|s\rangle$.  For $h=0$, the
ground state is the magnetized state $|\,-\,\rangle$.  The
lowest-energy excited states are kink-antikink states, like those
reported in Eq.~\eqref{kakstates} for small values of $g$. The energy
difference $\Delta(L,h)$ between the ground state and the
kink-antikink states is finite for $h=0$, but rapidly decreases
with increasing $h$, up to a pseudotransition point $h_{\rm tr}$ where
the state $|\,-\,\rangle$ and the lowest-energy kink-antikink state
have an avoided crossing.  The magnetic-field value $h_{\rm tr}$ can
be defined as the one for which the gap $\Delta(L,h)$ takes its
minimum value.  As discussed in Ref.~\cite{PRV-18-fowb}, $h_{\rm tr}$
scales as $1/L$, while $\Delta(L,h_{\rm tr})$ vanishes exponentially
with $L$.
%

\begin{figure}[!t]
  \includegraphics[width=0.47\textwidth]{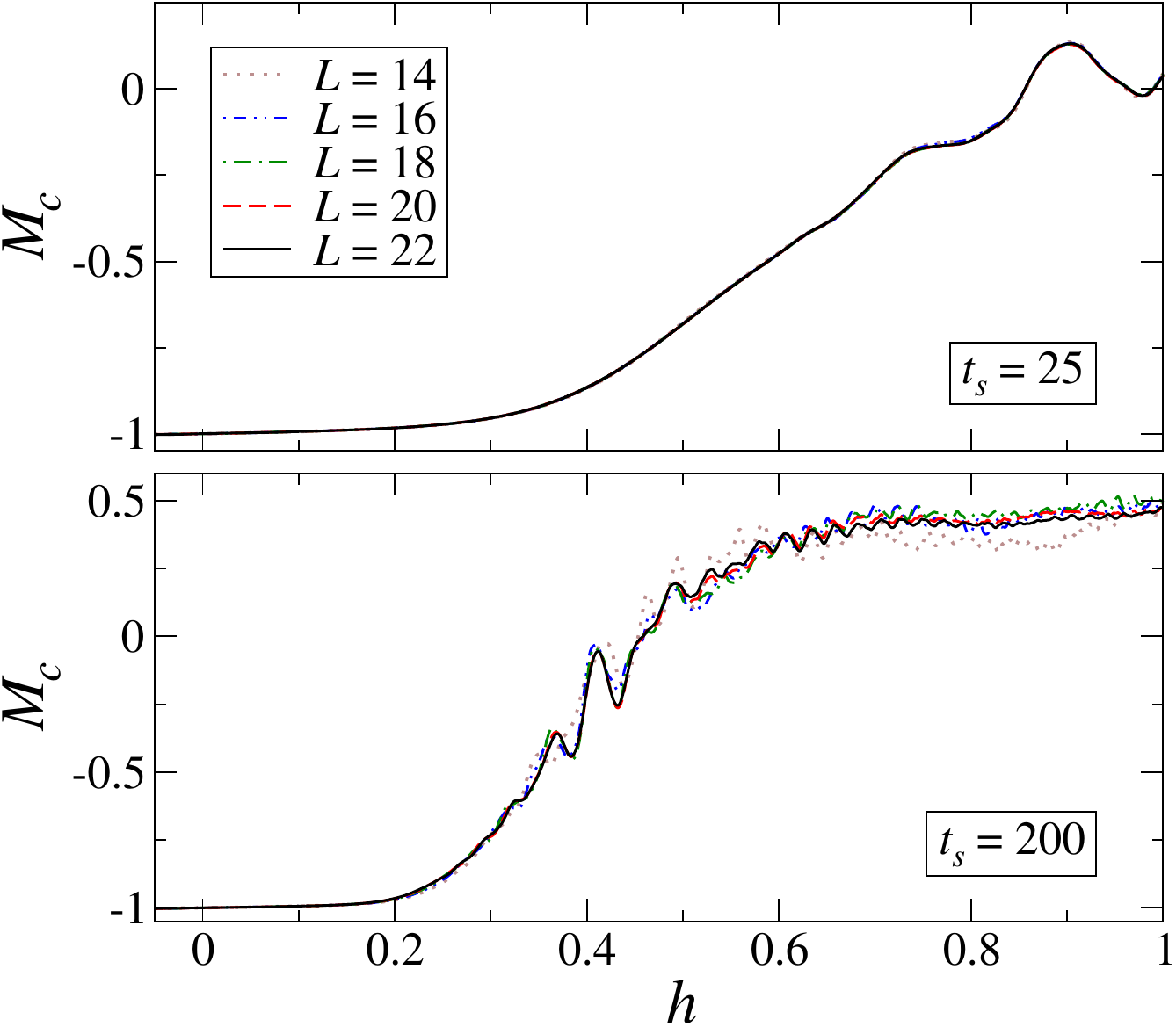}
  \caption{Rescaled central magnetization for $t_s = 25$ (top) and $200$
    (bottom) for EFBC. Results should be compared with those for OBC, 
    reported in Fig.~\ref{OBCinfL}. Note that convergence is faster in
    the EFBC case. Indeed, while EFBC data for $L=22$ and $t_s=200$
    are clearly asymptotic, in the OBC case significant finite-size
    effects are still present for $L=22$ when $t_s = 200$ (see the
    bottom panel of Fig.~\ref{OBCinfL}).  }
  \label{EFBCinfL}
\end{figure}

\begin{figure}[!t]
  \includegraphics[width=0.47\textwidth]{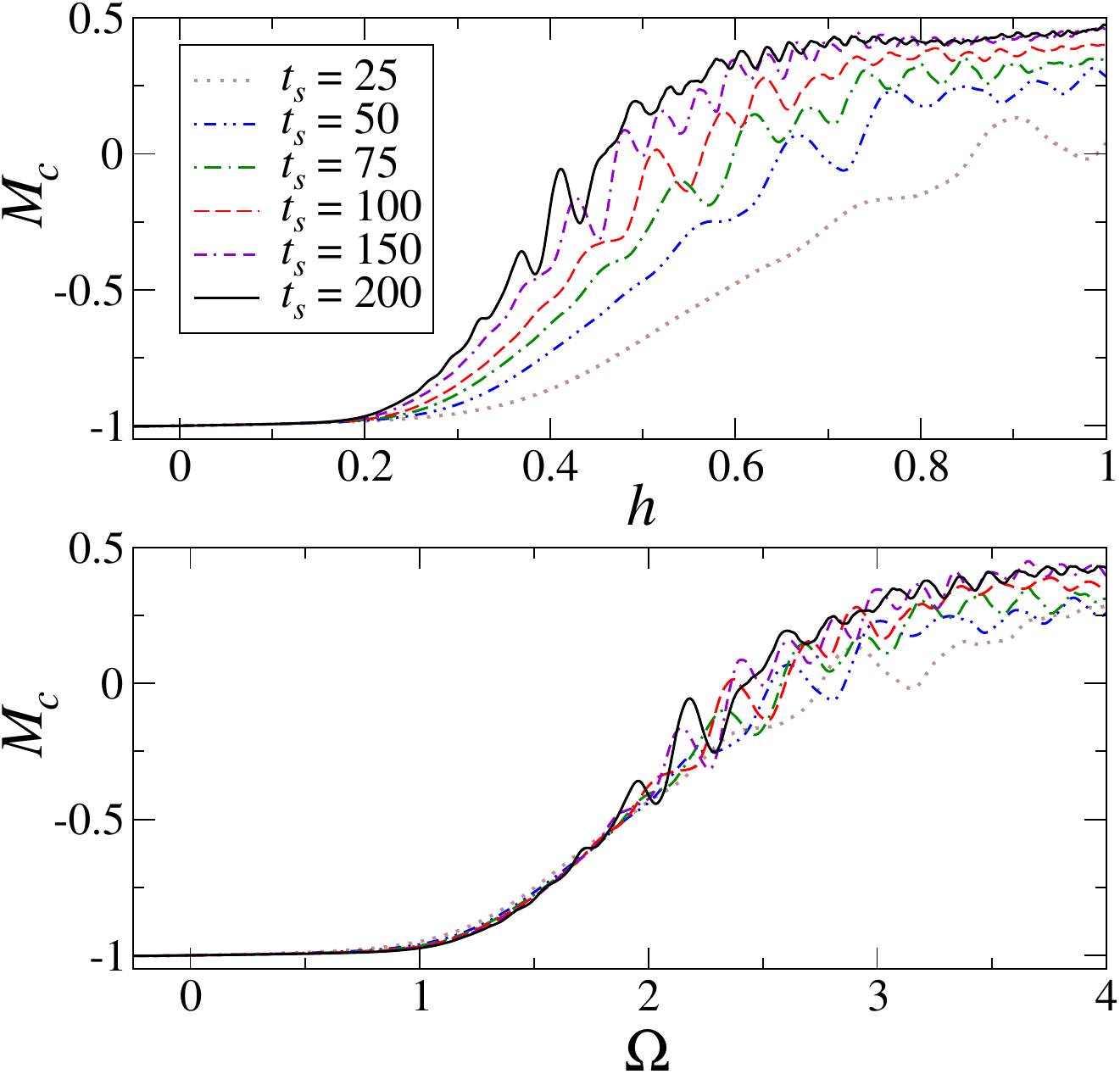}
  \caption{Infinite-size rescaled central magnetization for different
    values of $t_s$, vs $h(t)$ (top) and $\Omega(t)$ (bottom).
    The data have been obtained for EFBC. Results should be compared
    with those for OBC, reported in Fig.~\ref{OBCtssca}.
    Also for EFBC, we observe a reasonable scaling when the data are
    plotted as a function of the rescaled variable $\Omega(t)$ (bottom
    panel).  }
  \label{EFBCtssca}
\end{figure}

For $h\approx h_{\rm tr}$ one can define a static
FSS~\cite{PRV-18-fowb} and an OFSS~\cite{PRV-20}, as reviewed in
Sec.~\ref{ofss}.  Here we will not consider the OFSS behavior,
focusing instead on the scaling behavior of the KZ dynamics in the TL.
We follow the same procedure outlined for systems with OBC.  In
Figs.~\ref{EFBCinfL} and~\ref{EFBCtssca} we report results for the
central magnetization of Ising chains at $g=0.5$, which should be
compared with those reported in Figs.~\ref{OBCinfL}
and~\ref{OBCtssca}, which refer to chains with OBC.  These results
show that, although the OFSS behavior in systems with OBC and EFBC
differs, the behavior in the TL is substantially similar in the two
cases. Indeed, also for EFBC, the infinite-length data scale as a
function of $\Omega$ defined in Eq.~\eqref{omegadef}. Moreover, the
magnetization curves for chains with EFBC are (even quantitatively, as
we discuss in Sec.~\ref{compres}) very similar to those obtained for
OBC. Note, however, that EFBC data show a faster convergence to the
infinite-size limit behavior.  This is particularly evident for the
data with $t_s = 200$: The clearly visible size-dependent time
oscillations of the central magnetization for chains with OBC (bottom
panel of Fig.~\ref{OBCinfL}) are significantly smaller in systems with
EFBC (bottom panel of Fig.~\ref{EFBCinfL}).  This allows us to observe
a better collapse of the curves of the infinite-size magnetization,
especially in the range of $\Omega(t) \lesssim 2$ (see bottom panel of
Fig.~\ref{EFBCtssca}).

\subsection{Opposite fixed boundary conditions}
\label{ofbcsec}

Let us now discuss the KZ dynamics in systems with OFBC.  We consider a
chain of $L+2$ sites with Hamiltonian~\eqref{hedef} and we restrict
the state space to states $|s\rangle$ satisfying $\hat\sigma^{(1)}_0
|s\rangle = +|s\rangle$ and $\hat\sigma^{(1)}_{L+1} |s\rangle =
-|s\rangle$ (i.e., we fix opposite polarizations for the two boundary
spins at $x=0$ and $x=L+1$).  In this case, the low-energy spectrum is
characterized by single-kink states, as discussed in detail in
App.~\ref{AppOBC}, with energy gaps of order $1/L^2$. In particular,
the ground-state gap is~\cite{CJ-87,CPV-15,CPV-15b}
\begin{equation}
  \Delta(L) = {3g\over 1-g} {\pi^2\over L^2} + O(L^{-3}).
  \label{deltalofbc}
\end{equation}
The KZ dynamics of finite-size systems shows the OFSS behavior
outlined in Sec.~\ref{ofss}, as for systems with OBC.
However, the different size behavior of the gap and the presence of an
infinite tower of states that become degenerate for $L\to \infty$
leads to notable differences with respect to the OBC case. First,
scaling variables depend on powers of $L$; for instance, the relevant
time scale $T(L)=L/\Delta(L)^2$ scales as $T(L)\sim L^5$. Moreover,
scaling functions cannot be derived by using a two-level effective
model. On the other hand, one should include all single-kink states
that characterize the low-energy spectrum (this is done in
App.~\ref{AppOFSS}).

\begin{figure}[!t]
  \includegraphics[width=0.47\textwidth]{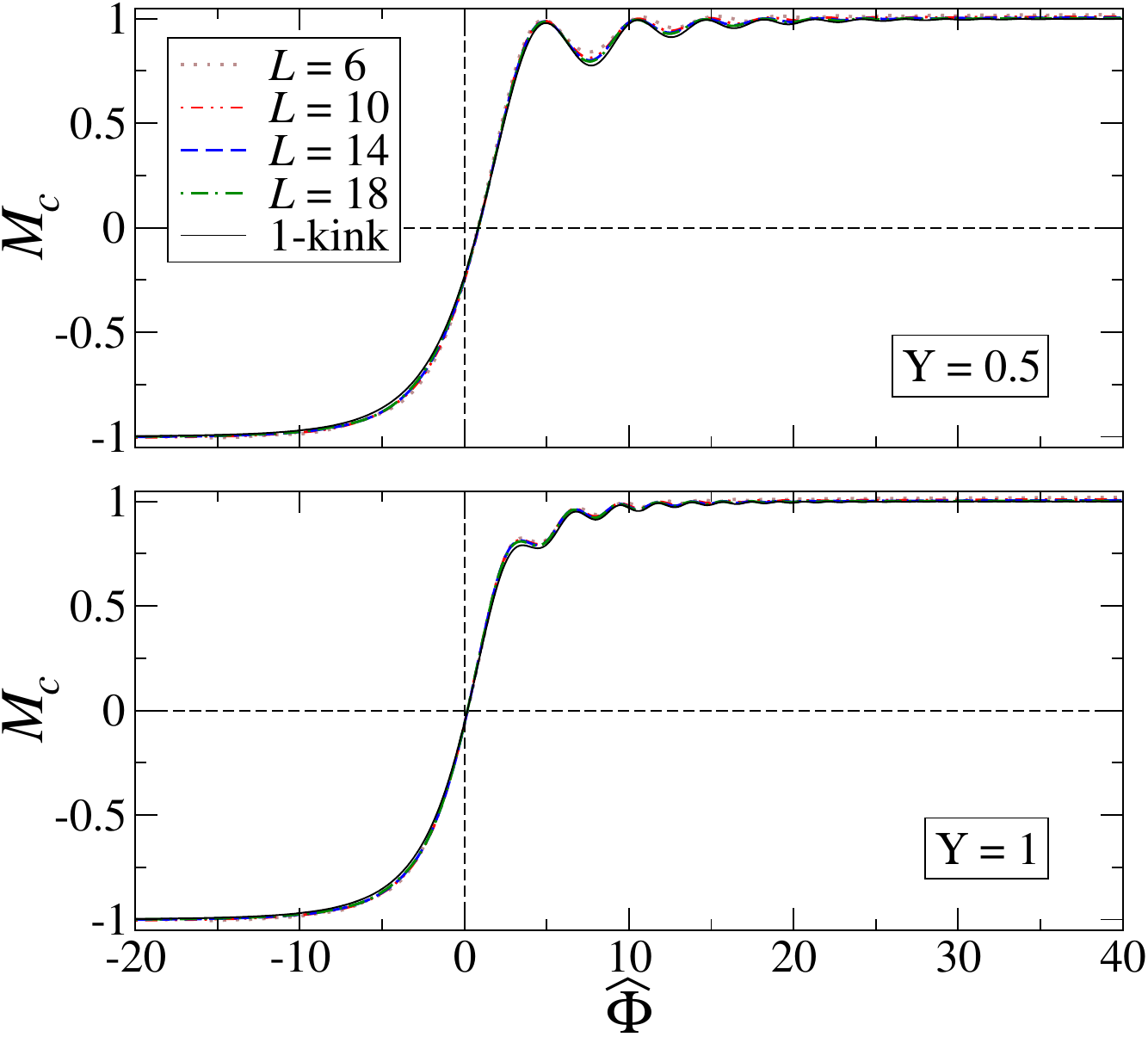}
  \caption{Rescaled central magnetization for a chain with OFBC, as a
    function of the rescaled variable $\widehat{\Phi}$, defined in
    Eq.~\eqref{katdef}.  We report results for two values of
    $\Upsilon$, defined in Eq.~(\ref{upsilondef}): $\Upsilon = 0.5$
    (top) and $1$ (bottom).  The colored dashed-dotted curves
    have been obtained by numerically solving the KZ dynamics in the
    full Hilbert space, for different system sizes (see legend).  The
    continuous black line (1-kink) is the result of a computation
    performed in the restricted single-kink model discussed in
    App.~\ref{AppOFSS}, with an appropriate rescaling of
    $\widehat{\Phi}$ and $\Upsilon$ (see text).  }
  \label{OFBCofss}
\end{figure}

As the OFSS regime in systems with OFBC has not yet been considered in
the literature, we briefly report some results. Figure~\ref{OFBCofss}
shows the time evolution of the central
magnetization defined in Eq.~(\ref{MMcdef}), for $L\le 18$ and $g=0.5$.
The data are plotted vs the rescaled variable $\widehat{\Phi}$,
defined in Eq.~(\ref{katdef}), and for two fixed values of
$\Upsilon$, defined in Eq.~(\ref{upsilondef}).  The data scale
as predicted by Eq.~\eqref{mtsl}, clearly supporting the OFSS theory
presented in Sec.~\ref{ofss}. We have also compared the numerical data
with those computed in a model restricted to states that are combinations of 
single-kink states. As discussed in App.~\ref{AppOFSS}, this
model becomes exact for $g\ll 1$ and $h\ll 1/L$. To compare the
single-kink results with the numerical ones computed in the full model
for $g=0.5$, one should keep into account that the normalization of
the scaling variables $\widehat{\Phi}$ and $\Upsilon$ is not
universal, so we must rescale $\widehat{\Phi}$ and $\Upsilon$ before
comparing results computed for different values of $g$. We thus
introduce two constants $c_{\scriptscriptstyle \Phi}$ and
$c_{\scriptscriptstyle \Upsilon}$ such that $\widehat{\Phi}_{\rm 1k} =
c_{\scriptscriptstyle \Phi} \widehat{\Phi}$, $\Upsilon_{\rm 1k} =
c_{\scriptscriptstyle \Upsilon} \Upsilon$, where the suffix ``1k"
specifies that the variables refer to the single-kink model. The
single-kink results are reported in Fig.~\ref{OFBCofss} for
$c_{\scriptscriptstyle \Upsilon} = 1/c_{\scriptscriptstyle \Phi} =
1.12$.  The scaling is excellent, confirming that the restricted model
effectively encodes the scaling features of the dynamics in the OFSS
limit.  We have also verified the consistency of the single-kink
approach, analyzing the average magnetization. As expected, numerical
data for $g=0.5$ and single-kink data approach the same scaling curve
(data not shown) if we fix $c_{\scriptscriptstyle \Phi}$ and
$c_{\scriptscriptstyle \Upsilon}$ to the same values obtained from the
analysis of the central magnetization.  As a final remark, note that
the estimates of the nonuniversal rescalings are such that the scaling
variable $\Theta = t \, \Delta(L) = \widehat{\Phi} \, \Upsilon$ (see
Sec.~\ref{ofss}) does not require any rescaling.

It is also interesting to discuss the behavior of the OFSS scaling
functions for $t_s \ll T(L)$. If the two-level approximation works
(this is the case of PBC and OBC), we can use Eq.~(\ref{mofss}) to
predict the large-time behavior of the scaling functions, obtaining
\begin{equation}
  {\cal M}(\Upsilon,\widehat\Phi\to\infty) = 1 - 2 \, e^{- \pi \Upsilon/2}.
  \label{limitLZ}
\end{equation}
For $t_s \ll T(L) $, we have $\Upsilon \to 0$, implying that the
rescaled magnetization does not change in the KZ dynamics. A similar
result is obtained for OFBC. To study the OFSS behavior for $t_s \ll
T(L)$, we define the $L$-independent scaling variable
\begin{equation}
  W = \widehat{\Phi} \, \Upsilon^{3/5} \sim t/t_s^{2/5},
  \label{Wdef}
\end{equation}
and rewrite the scaling relation~\eqref{mtsl} as 
\begin{equation}
  M(t,t_s,h_i,L) \approx \widetilde{\cal M}(\Upsilon, W).
  \label{mtsl2}
\end{equation}
The limit $\Upsilon\to 0$ at fixed $W$ is discussed in
App.~\ref{AppOFSS}, finding
\begin{equation}
  \widetilde{\cal M}(\Upsilon, W) = - 1 +
  O(\Upsilon^{1/5}).
  \label{mtildesca}
\end{equation}
Also in this case, the OFSS scaling function is trivial for $t_s \ll T(L)$.

\begin{figure}[!t]
  \includegraphics[width=0.47\textwidth]{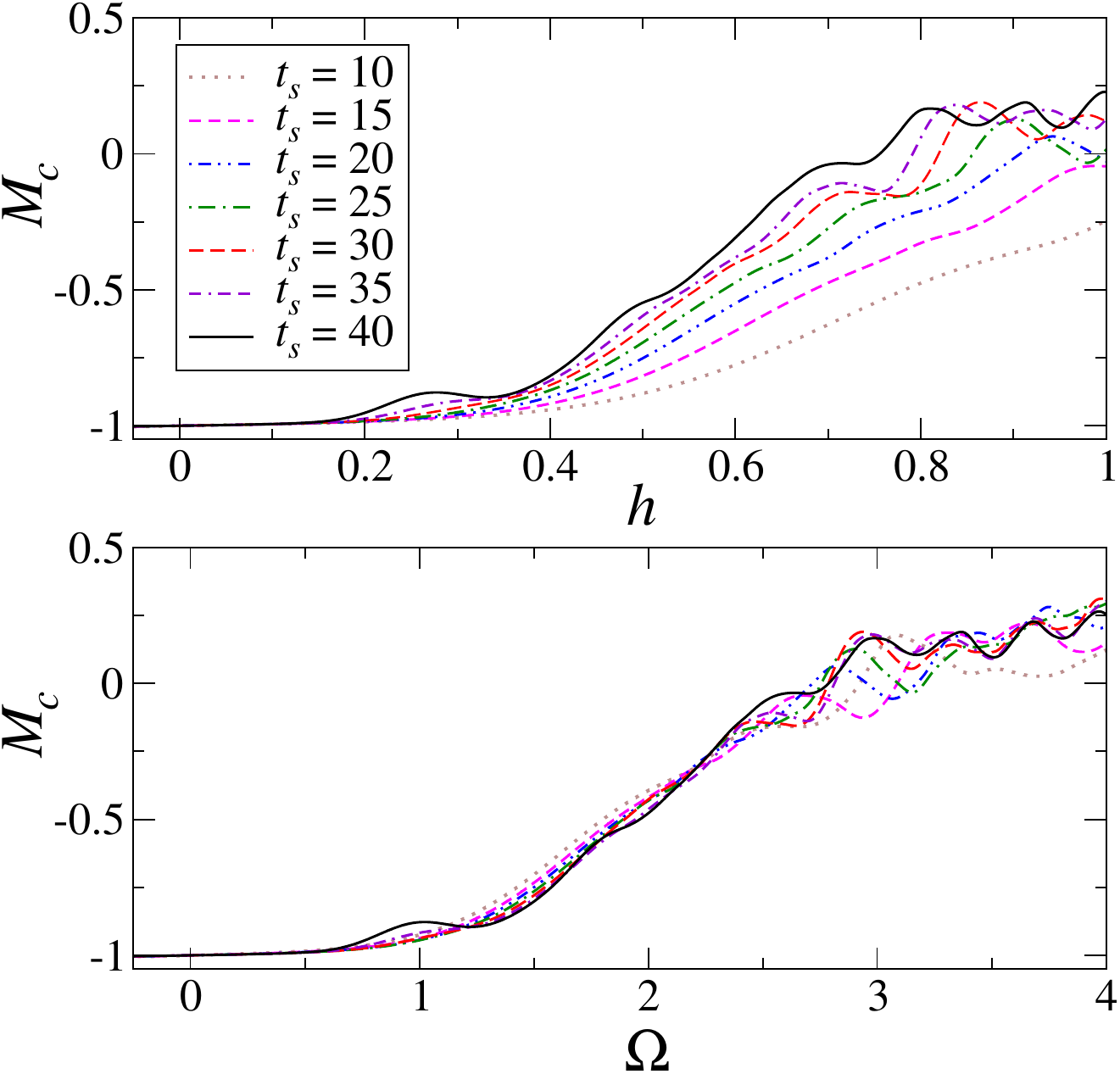}
  \caption{Infinite-size KZ dynamics for chains with OFBC: results for
    the rescaled central magnetization vs $h$ (top) and $\Omega$ (bottom).
    Figs.~\ref{OBCtssca} and~\ref{EFBCtssca} report analogous data for
    systems with OBC and EFBC, respectively.  We display data only for
    $t_s \leq 40$, since data for larger values of $t_s$ 
    are affected by large size corrections.
    As in the OBC and EFBC cases,
    we observe a fairly good collapse of the magnetization curves, when
    plotted vs the rescaled variable $\Omega(t)$ (bottom panel).}
  \label{OFBCinfL}
\end{figure}

Finally, we study the TL of the KZ dynamics. We proceed as we have
already done for systems with other BC (see Sec.~\ref{obcsec} for OBC
and Sec~\ref{EFBCsec} for EFBC).  For chains of size $L\le 22$---those
for which the dynamics can be studied in a reasonable amount of
time---we observe significant time oscillations for $t_s\gtrsim 50$.
For $h\approx 0$, the single-kink reduced model can be used to
understand their nature (see App.~\ref{AppOFSS}): It turns out that
they are due to the large number of quasidegenerate states (they are
not present for OBC or PBC where only two states are relevant for
$h\approx 0$) with $M\approx -1$, which have a significant overlap
with the system state $|\Psi(t)\rangle$, as soon as $t > 0$.
Oscillations decrease with the system size as $1/L$, but increase
with the time scale as $t_s^{1/2}$. For this reason, we are only
able to estimate the infinite-size evolution for $t_s\le 40$.  The
numerical results for the rescaled central magnetization are reported
in Fig.~\ref{OFBCinfL} as a function of the field $h(t)$ (top) and of
$\Omega(t)$ (bottom).  Finite-size corrections appear to be under
control for chains of length $L=22$.  When plotted vs the rescaled
variable $\Omega(t)$, the central magnetization curves show a
reasonably $t_s$-independent scaling behavior, at least up to $\Omega
\approx 2$. For larger values, although oscillations are clearly
present, all curves have apparently the same average behavior.

\subsection{Antiperiodic boundary conditions}
\label{abcsec}

\begin{figure}[!t]
  \includegraphics[width=0.47\textwidth]{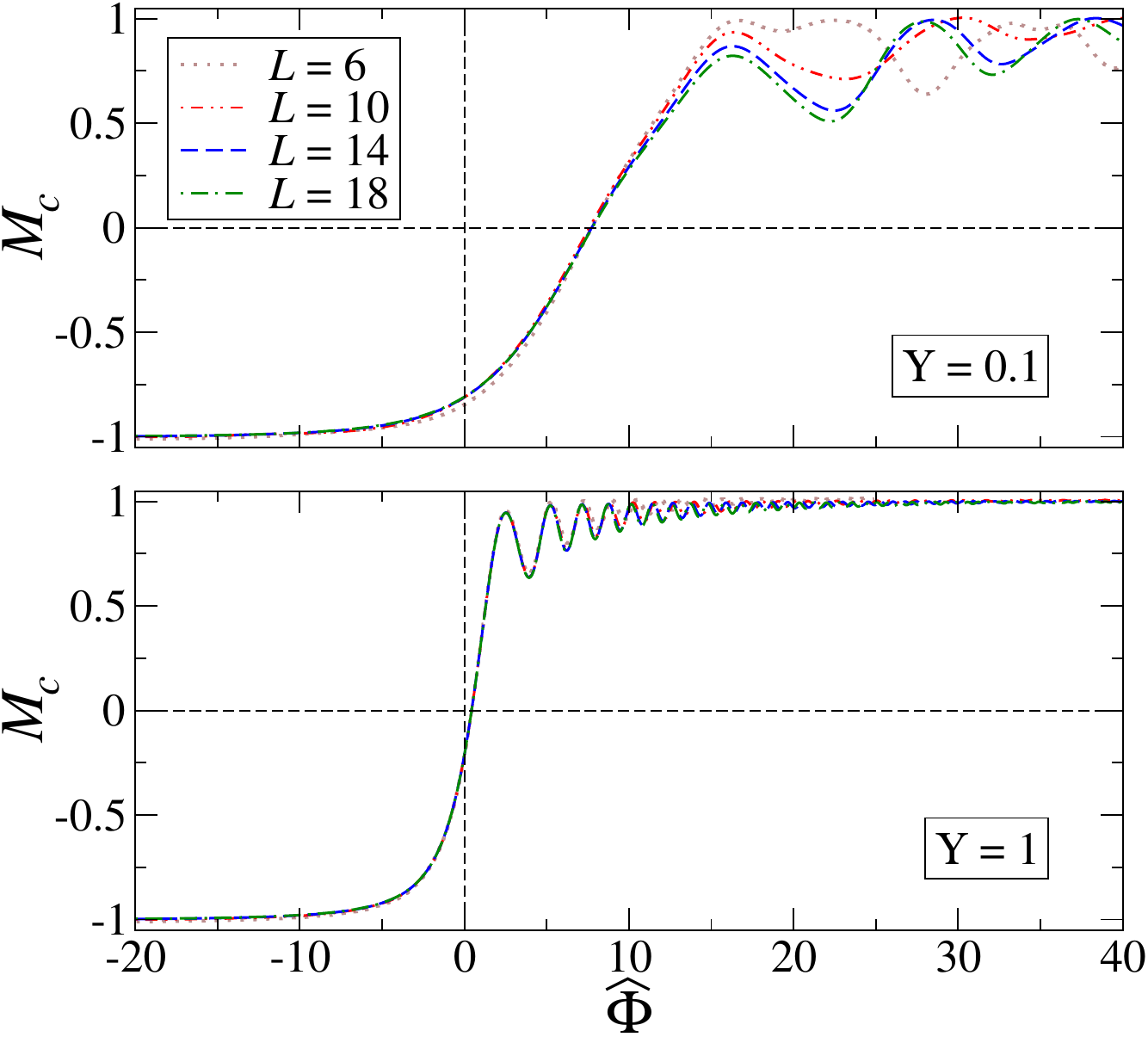}
  \caption{Rescaled central magnetization for systems with ABC, for 
    two different values of $\Upsilon= 0.1$ (top) and $1$ (bottom).
    Figure~\ref{OFBCofss} reports the same quantity for OFBC.}
  \label{ABCofss}
\end{figure}

\begin{figure}[!b]
  \includegraphics[width=0.47\textwidth]{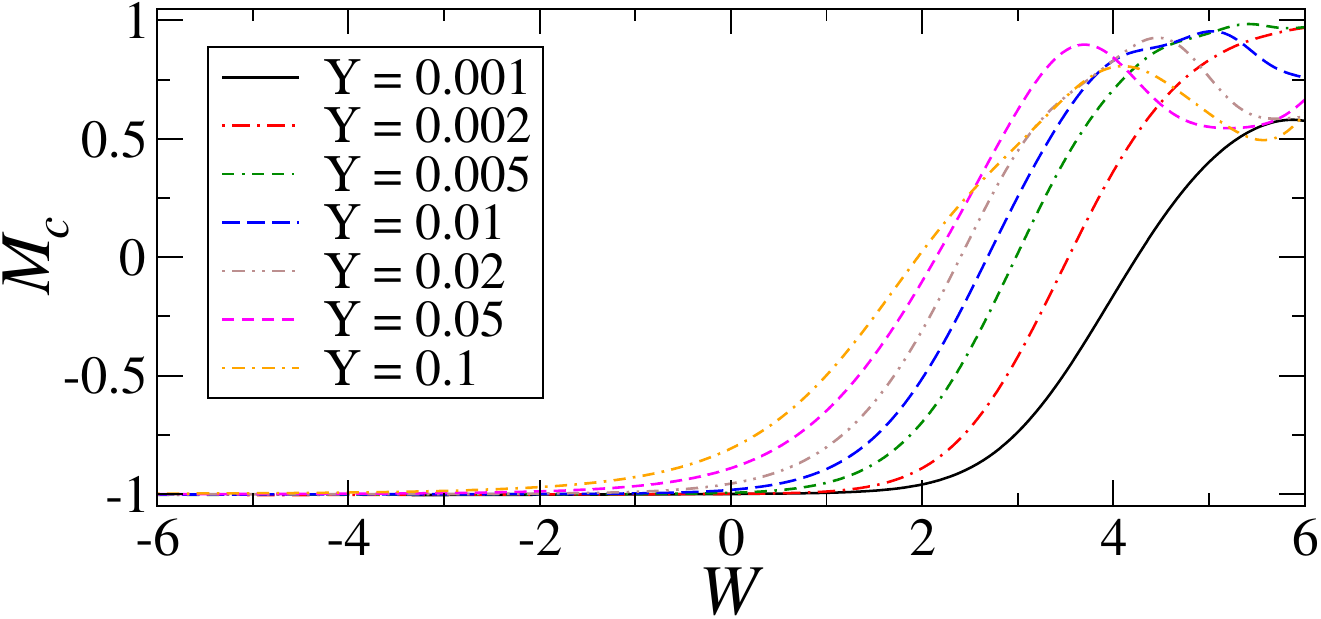}
  \caption{Rescaled central magnetization $M_c$ defined in
    Eq.~\eqref{MMcdef}, for a chain with ABC, as a function of the
    scaling variable $W = \widehat{\Phi} \, \Upsilon^{3/5}$ defined in
    Eq.~\eqref{Wdef}.  The curves refer to different values of
    $\Upsilon$, as indicated in the legend, for the largest available
    sizes ($L=22$ for $\Upsilon \leq 0.02$, $L=20$ for $\Upsilon \geq
    0.05$).}
  \label{ABCfsslogsing}
\end{figure}

We now consider the KZ dynamics in systems with ABC. Their low-energy
spectrum is similar to that of systems with OFBC: there is an infinite
tower of states with energy gaps of order $1/L^2$ that become
degenerate in the infinite-volume limit. In particular, the
ground-state energy gap is~\cite{CJ-87,CPV-15,CPV-15b}
\begin{equation}
  \Delta(L) = {g\over 1-g} {\pi^2\over L^2} + O(L^{-4}).
  \label{deltalabc}
\end{equation}
Again, we have verified the presence of an OFSS regime for $h\approx 0$.
In Fig.~\ref{ABCofss} we show the evolution of the central
magnetization, as obtained from numerical simulations of chains with
$L\le 18$ sites at $g=0.5$.  As occurring in the OFBC case, our results
nicely agree with the OFSS relation~\eqref{mtsl}.  For $t_s \ll T(L)$,
we expect the OFSS behavior to be trivial, with $M_c \approx -1$, with
corrections that should scale as $\Upsilon^{1/5}$ at fixed $W$ [$W$ is
  defined in Eq.~(\ref{Wdef})], as in the OFBC case. To verify this
prediction, in Fig.~\ref{ABCfsslogsing}, we show the magnetization as
a function of $W$, for different values of $\Upsilon$. At fixed $W$,
the central magnetization decreases for $\Upsilon \to 0$ towards $-1$
(all spins remain frozen in the fully polarized state), as expected.

\begin{figure}[!t]
    \includegraphics[width=0.47\textwidth]{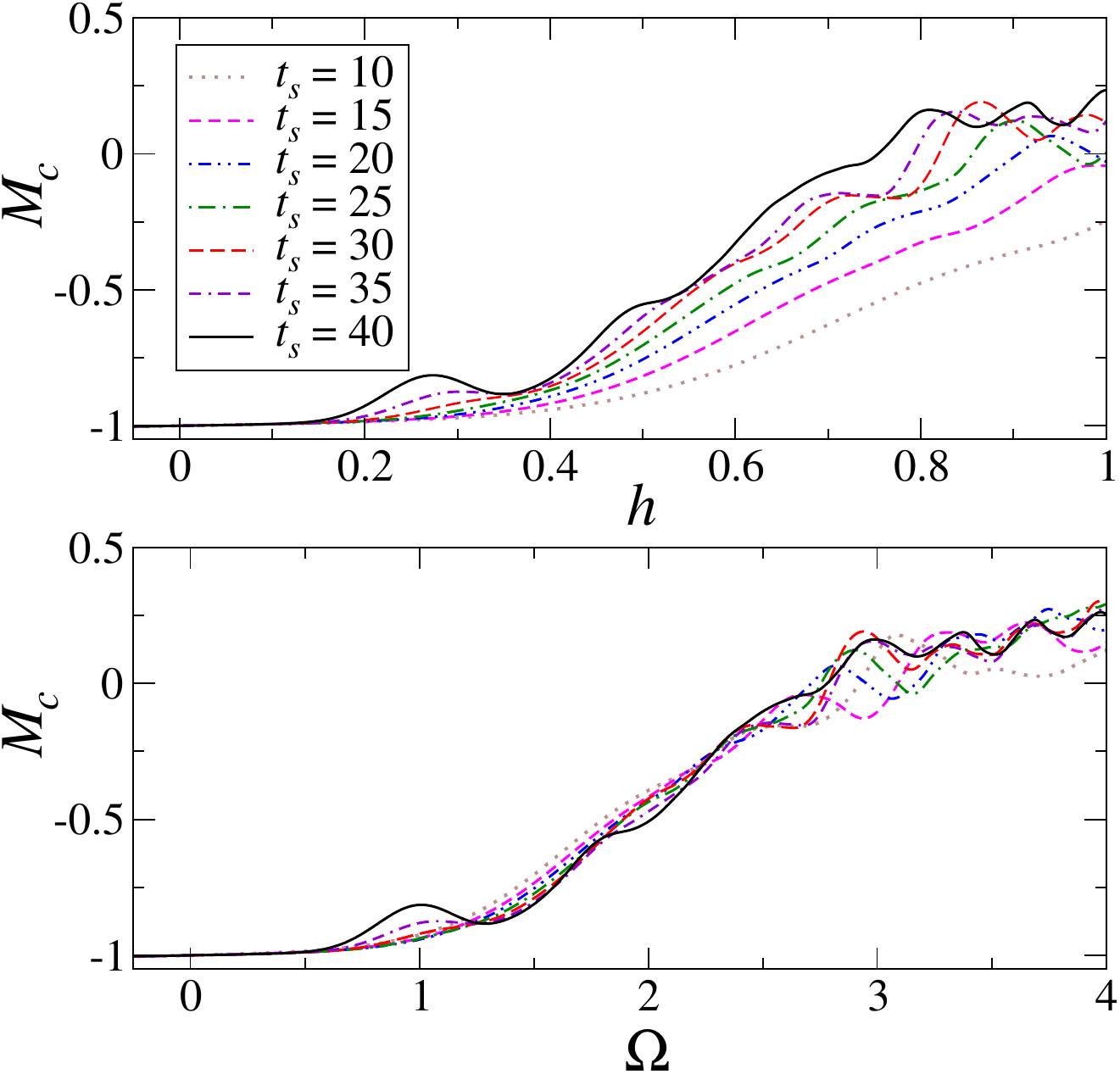}
    \caption{Infinite-size KZ dynamics for chains with ABC: results for
      the rescaled central magnetization vs $h$ (top) and $\Omega$ (bottom) for
      systems of size $L=22$.  Figures~\ref{OBCtssca},~\ref{EFBCtssca},
      and~\ref{OFBCinfL}, report analogous data for systems with OBC,
      EFBC, and OFBC, respectively.}
    \label{ABCtssca}
\end{figure}

Finally, we consider the dynamics in the TL, adopting the same
protocol already used for other types of BC in Secs.~\ref{obcsec},
\ref{EFBCsec}, and~\ref{ofbcsec}.  The top panel of
Fig.~\ref{ABCtssca} shows the central-magnetization curves vs the
field $h(t)$, for the values of $t_s$ for which we obtained a
reasonable approximation of the infinite-size limit.  Note that, for a
given values of $t_s$, the TL in ABC systems is observed for sizes
significantly larger than those needed when considering PBC, OBC, and
EFBC. Thus, we are only able to obtain infinite-size data for small
values of $t_s$. The behavior is very similar to that observed in the
OFBC case, probably as a consequence of the analogous nature of the
spectrum, characterized by an infinite number of degenerate states for
$L\to\infty$ and $h = 0$.  In particular, the numerical outcomes
follow, even quantitatively, those obtained with OFBC and reported in
Fig.~\ref{OFBCinfL}, although with tiny differences that shift
convergence to even larger values of $L$. For this reason, we can reliably
obtain infinite-size results only for $10 \leq t_s \leq 40$, with systems
of size $L\le 22$.  Nonetheless, we are able to find a reasonable
evidence that the infinite-size KZ dynamics is characterized by a
scaling in terms of the variable $\Omega=t/\tau_s$, with
$\tau_s=t_s/\ln t_s$.  Indeed, data for different $t_s$ apparently fall
onto a single curve when plotted in terms $\Omega$
(bottom panel of Fig.~\ref{ABCtssca}).

\subsection{Finite-$t_s$ corrections in the thermodynamic limit: an 
analysis for PBC systems} 

We have phenomenologically addressed the existence of a well-defined
TL for the dynamics at fixed $t_s$. Then, we have somehow verified the
existence of a scaling behavior as a function of the variable $\Omega$
for large enough values of $t_s$.  However, to achieve a good control of
the large-$t_s$ limit, it is also important to understand the nature
of the finite-$t_s$ corrections.  We perform this analysis in systems
with PBC, since in this case we can obtain results for longer chains
(up to $L=26$) exploiting translation invariance.

\begin{figure}[!t]
  \includegraphics[width=0.48\textwidth]{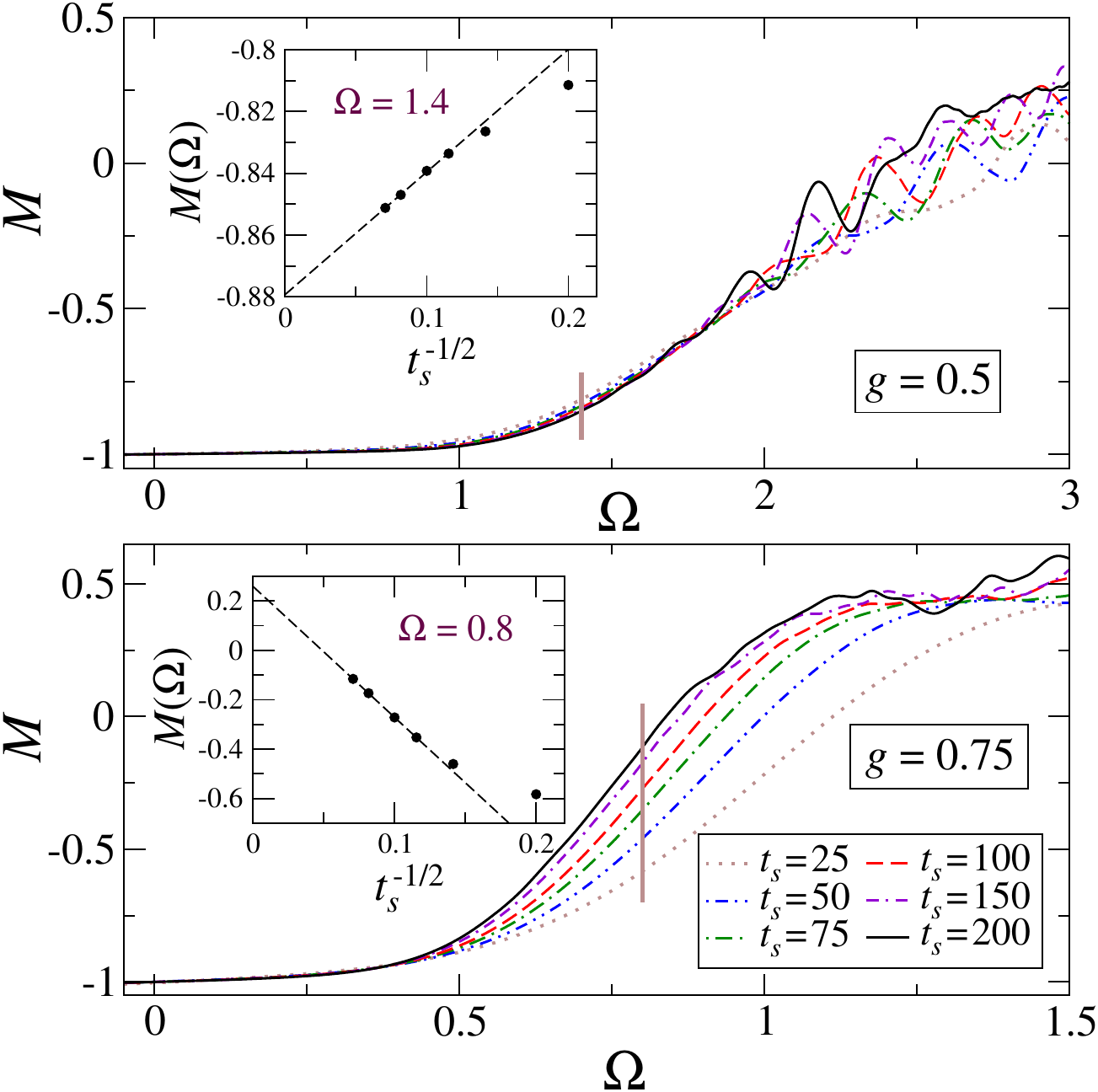}
  \caption{Main frames: The rescaled longitudinal magnetization $M$,
    for chains with PBC and $L=26$ sites, vs the rescaled variable
    $\Omega(t) = h(t) \ln t_s$. The curves correspond to different
    values of $t_s$ (see legend), up to the largest value for which we
    verified the convergence to the infinite-size limit. The top
    panel is for $g = 0.5$, while the bottom panel is for $g = 0.75$.
    Insets: $M$ at fixed $\Omega$, plotted vs
    $t_s^{-1/2}$. The value of $\Omega$ has been chosen in the region
    in which the magnetization is growing faster ($\Omega = 1.4$ for
    $g=0.5$, $\Omega = 0.8$ for $g=0.75$).  The dashed lines are fits
    to the numerical data (circles) for the largest values of $t_s$.}
  \label{PBCtssca}
\end{figure}

Figure~\ref{PBCtssca} reports the average magnetization data vs
$\Omega$ for two different values of $g=0.5$ (top) and $g=0.75$
(bottom). To understand the corrections, we select two different
values of $\Omega$ and study the behavior of $M$ at fixed $\Omega$
(vertical bold lines) as a function of $t_s$.  The results are shown
in the corresponding insets. In both cases, our numerical data lie
on a straight line, when plotted vs $t_s^{-1/2}$,
indicating that corrections scale as $t_s^{-1/2}$.

\subsection{Comparison of the results in the thermodynamic limit}
\label{compres}

We finally compare the results for the central magnetization obtained
in the infinite-size limit for different BC. We consider the OBC,
EFBC, OFBC, ABC and PBC results presented above.  This is a nontrivial
comparison, because the main features of the scaling behavior at a
FOQT generally depend on the BC. For instance, the size dependence of
the OFSS scaling variables as well as the corresponding scaling
functions depend on the BC. Here, we wish to understand whether the
dynamic scaling behavior in the TL depends or not on the BC.

\begin{figure}[!t]
  \includegraphics[width=0.47\textwidth]{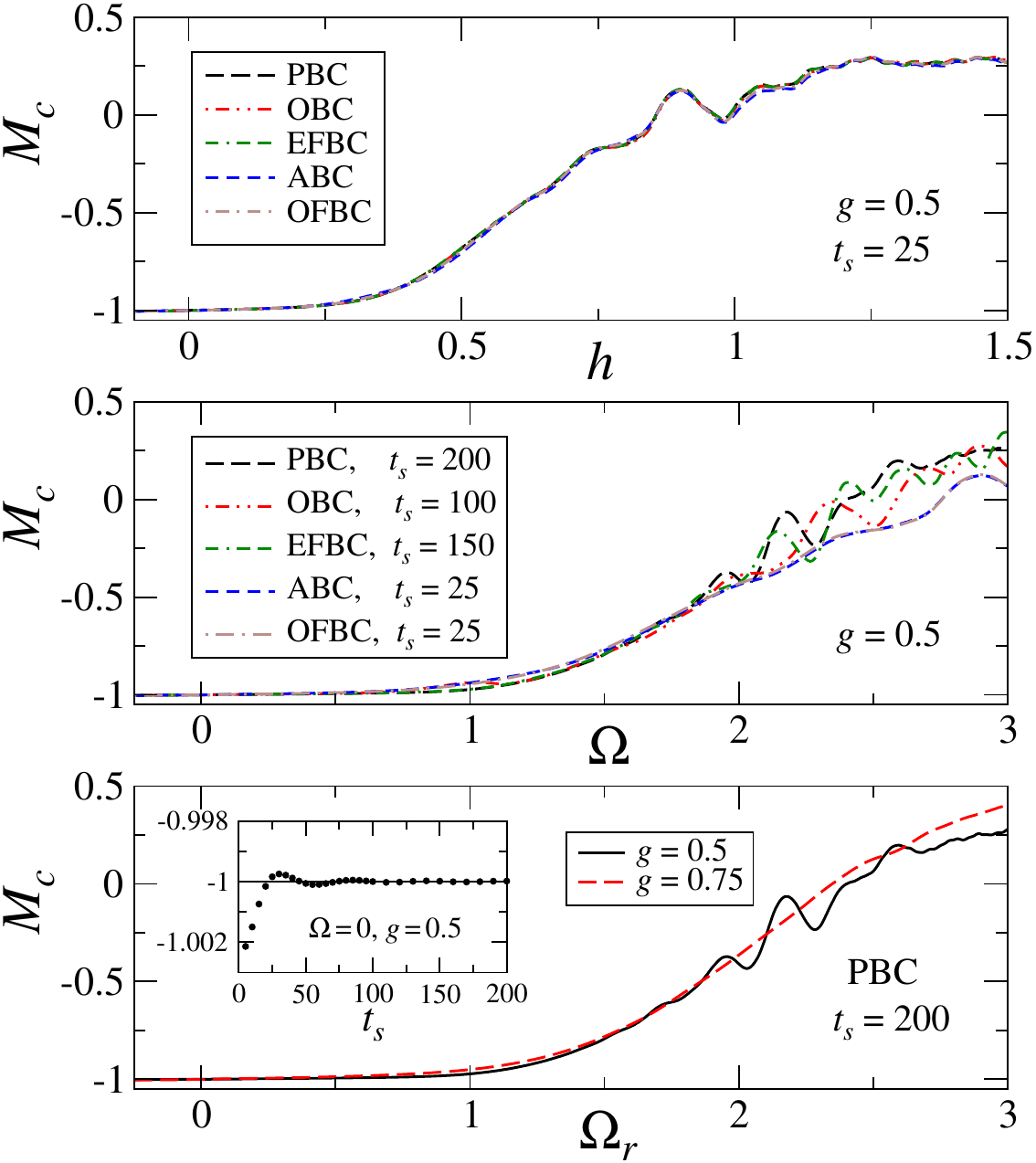}
  \caption{Comparison of the KZ dynamics in the TL: We report results
    for the rescaled central magnetization $M_c$ for different BC and
    transverse fields $g$.  Top: $M_c$ as a function of $h(t)$, at
    fixed $t_s=25$ and $g=0.5$, for different BC.  Middle: $M_c$ as a
    function of $\Omega(t) = h(t)\ln t_s$ for $g=0.5$ and different
    BC. For each type of BC, we consider the largest value of $t_s$
    for which we have a robust evidence that the infinite-size limit
    has been reached.  Bottom: $M_c$ for two different values of $g$,
    for $t_s=200$ and PBC, as a function of the rescaled variable
    $\Omega_r = c_{\scriptscriptstyle \Omega} \, \Omega$;
    $c_{\scriptscriptstyle \Omega} = 1$ for $g=0.5$ and
    $c_{\scriptscriptstyle \Omega} = 2.8$ for $g=0.75$.  The inset shows
    $M_c$ vs $t_s$, at $\Omega=0$, $g=0.5$, and PBC. The numerical
    data shown in this figure have been obtained for chains of length
    $L=22$ for all types of BC except PBC; in the PBC case we consider
    chains with $L=26$ sites.}
  \label{TLcomp}
\end{figure}

In the top panel of Fig.~\ref{TLcomp}, we report the evolution of the
infinite-size central magnetization for systems with PBC, OBC, EFBC,
ABC, and OFBC, for $t_s = 25$. The data indicate that the
infinite-size evolution at fixed $t_s$ is the same for all BC.  The
apparent tiny differences between the various curves can be
explained by size corrections that are present for finite values of
$L$ (we report results for $L=26$ for PBC and for $L=22$ for all the
other BC).  A similar agreement can also be observed in the central
panel, where we plot the large-$t_s$ behavior, which is numerically
estimated by increasing $t_s$ until an acceptable convergence is observed
as a function of the rescaled variable $\Omega = h(t) \, \ln t_s$.
The agreement between the curves for different BC is good at least
for $\Omega \lesssim 2$, (for larger values oscillations set in).
This occurs despite the largest values of $t_s$ for which we can
observe the infinite-size evolution vary by one order of magnitude (as
already discussed, ABC and OFBC display large finite-size corrections,
which limit us to study $t_s \approx 25$).

We also checked the universality of the magnetization dynamics with
respect to the system parameters, comparing the evolution for two
different values of the transverse field: $g=0.5$ and $g=0.75$.
For this purpose, we have considered PBC, since translational invariance
allows us to consider larger sizes and, therefore, to obtain results
with smaller finite-size corrections. To perform a correct comparison,
we should however take into account that $\Omega$ is defined up to a
nonuniversal normalization. Therefore, we define the rescaled variable
$\Omega_r = c_{\scriptscriptstyle \Omega} \Omega$, where
$c_{\scriptscriptstyle \Omega}$ is a $g$-dependent constant that can
be tuned to obtain a universal (i.e., $g$-independent) scaling
behavior.  Results are shown in the bottom panel of Fig.~\ref{TLcomp}.
We observe a reasonable agreement up to $\Omega_r \approx 2.5$, which
is the interval of values of $\Omega_r$ in which we have a robust
evidence that the TL has been reached.

In conclusion, although we could perform computations up to moderately
large sizes, we believe that our numerical results for the KZ dynamics
provide a robust evidence of the existence of a well defined TL for
fixed values of $t_s$, which is independent of the BC.  Moreover, the
infinite-size evolution for large values of $t_s$ is apparently
characterized by a scaling behavior in terms of $\Omega=t/\tau_s$,
with $\tau_s=t_s/\ln t_s$.

We stress that the scaling behavior of the KZ dynamics in the TL is
distinct from, and substantially unrelated with, the OFSS behavior,
which occurs in a tiny interval of width $\Delta(L)/L$ [$\Delta(L)$ is
  the ground-state gap for $h=0$] around the transition point $h =
0$. In terms of $\Omega$, the OFSS behavior would correspond to the
limit $\Omega\to 0$, and thus consistency of the OFSS and TL behavior
only leads to the trivial prediction $M = M_c = -1$ for $\Omega =
0$. This is confirmed by the data shown in the inset of the lower
panel of Fig.~\ref{TLcomp}, which shows $M_c$ at $t=0$ (thus
$\Omega=0$), with increasing $t_s$, for an Ising chain with PBC and
$g=0.5$. One can clearly observe a rapid convergence with $t_s$ of the
numerical data (circles) to the expected value $M_c=-1$ (straight
horizontal line).

We finally note that the observed behavior of the KZ
dynamics in the TL resembles the behavior occurring at a classical
spinodal point predicted in mean-field analyses of 
first-order transitions~\cite{Binder-87}. Indeed,
the quantum many-body system remains in the negative magnetization
$M=-1$ state up to $h^*>0$ (thus $t>0$), but with $h^*$ decreasing
logarithmically in the large-$t_s$ limit, as $h^* \sim 1/\ln t_s$.

\section{Conclusions}
\label{conclu}

This paper is a follow up of the analysis of the out-of-equilibrium KZ
dynamics at FOQTs presented in Ref.~\cite{PRV-25}. We consider the
one-dimensional quantum Ising chain in a transverse field $g$,
described by the Hamiltonian (\ref{hedef}). We discuss a KZ dynamics
with a time-dependent longitudinal field $h(t)=t/t_s$ that drives the
system across the {\em magnetic} FOQT at $h=0$ occurring for $|g|<J$.
We considerably extend the analysis reported in Ref.~\cite{PRV-25},
which was limited to quantum Ising chains with PBC. Here, besides PBC,
we also consider OBC, EFBC, OFBC, and ABC. For each type of BC, we
discuss the dynamic KZ behavior in the TL, which is obtained by taking
the infinite-size limit $L\to \infty$ at fixed $t_s$. Our novel
results provide a more complete and firmer characterization of the
emerging KZ scaling behavior in the TL.

We should stress that the scaling behavior emerging in
the TL is unrelated with the OFSS behaviors, which are observed in
finite systems, in a small interval of longitudinal fields around
$h\approx 0$, and which depend on the BC (here we report
additional results for OFBC and ABC). This is due to the fact that the
relevant states of the KZ dynamics in the TL are multi-kink states,
which are instead irrelevant in the OFSS
regime~\cite{PRV-18,PRV-20,RV-21,PV-24,PRV-25}.  This leaves open the
possibility of a unique TL of the KZ dynamics, independent of the BC.

Our numerical analyses confirm the emergence of a quantum
spinodal-like behavior of the KZ dynamics in the TL keeping $t_s$
fixed: the change of the magnetization from the initial value $m<0$ to
positive values $m>0$ occurs at positive values of $h=h_\star>0$ that
decrease as $h_\star \sim 1/\ln t_s$. Moreover, in the large-$t_s$
limit, the time evolution of the observables shows a universal scaling
behavior in terms of $\Omega = t/\tau_s$, where $\tau_s = t_s/\ln
t_s$, with $O(1/\sqrt{t_s})$ corrections.  These features of the KZ
dynamics are independent of the BC, confirming the existence of a well
defined TL, irrespective of the BC.  This should be somehow related to
fact that the length scale of the fixed-time connected correlation
functions is always bounded for any finite $t_s$ along the whole KZ
evolution in the TL (unlike the equilibrium behavior at the transition
point, which crucially depends on the BC).

We admit that we do not have a solid explanation for the
phenomenological observation that the KZ dynamics in the TL develops a
logarithmic spinodal-like behavior, which is only based on the
numerical evidence of our analyses. Of course, we cannot rule out
alternative, quantitatively similar, behaviors. For example, we cannot
exclude scaling in terms of a variable $\Omega$ in which $\ln t_s$ is
replaced by a power of $t_s$ with a small exponent.  However, we favor
a logarithmic spinodal-lile behavior, also because analogous
features are observed for the KZ dynamics in classical
systems, for instance, in classical two-dimensional and
three-dimensional Ising models driven across the magnetic first-order
transition line by a relaxational dynamics~\cite{PV-17,PV-25}.

Further investigations are needed to improve our understanding of this
spinodal-like behavior.  On one side, it would be worth identifying
simplified models that are able to explain and predict the observed
behavior. On the other side, it is important to improve the quality of
the numerical results, which is mainly limited by the chain sizes that
can be considered.  The present results have been obtained using
exact-diagonalization techniques, which are constrained to moderately
large chains with a few dozen of sites. Alternative approaches, such
as time-dependent DMRG methods~\cite{Schollwock-05, Schollwock-11},
may overcome this issue by truncating the effective Hilbert space of
the system. Such truncation is known to become efficient at low
energies, i.e., when the bipartite entanglement entropy satisfies an
area-law scaling.  Unfortunately, the regime we are interested in is
far from the adiabatic limit and a tower of excited states become
equally important for the analysis of the KZ dynamics across FOQTs in
the TL.
It would be also tempting to extend our analysis to other FOQTs, for
example in higher dimensions, both in finite-size systems and in the
TL.  Another interesting issue is related to the role of dissipation,
which can be introduced, e.g., in a Lindblad
framework~\cite{RV-21,DRV-20}.

We remark once again that the results presented in this work have been
observed in relatively small, or moderately large, systems ($L \approx
20$).  Therefore, given the need for high accuracy without necessarily
reaching scalability to large sizes, we believe that our predictions
may be checked experimentally, using, for instance, ultracold atoms in
optical lattices~\cite{Bloch-08, Simon-etal-11}, trapped
ions~\cite{Edwards-etal-10, Islam-etal-11, LMD-11, Kim-etal-11,
  Richerme-etal-14, Jurcevic-etal-14, Debnath-etal-16}, as well as
Rydberg atoms in arrays of optical microtraps~\cite{Labuhn-etal-16,
  Guardado-etal-18, Keesling-etal-19, BL-20} or even quantum computing
architectures based on superconducting qubits~\cite{Barends-etal-16,
  Gong-etal-16, CerveraLierta-18, Ali-etal-24}.  We also mention that
some recent experiments have already addressed the dynamics and the
excitation spectrum of quantum Ising-like chains~\cite{Gong-etal-16,
  LTDR-24, DMEFY-24}, thus opening possible avenues where the
envisioned behaviors at FOQTs can be observed in the near future.

\appendix

\section{FSS scaling behavior of the one-kink spectrum}
\label{AppOBC}

In this appendix we discuss the low-energy spectrum of a quantum Ising
chain of length $L$, for OFBC and OBC and small values of $h$,
extending the results of Ref.~\cite{PRV-25} for PBC.  We first focus
on OFBC.  Following the notation of Sec.~\ref{ofbcsec}, we consider a
chain of length $L_t = L + 2$ (sites are labelled with an integer $x$
running from 0 to $L+1$) and we restrict the Hilbert space to states
that are eigenstates of $\hat{\sigma}^{(1)}_0$ and
$\hat{\sigma}^{(1)}_{L+1}$ with eigenvalue $1$ and $-1$, respectively.
We consider small values of $g$ and $hL$, in such a way that we can
use perturbation theory to determine the energy levels.  For this
purpose, it is enough to consider the set of one-kink states that
become degenerate for $g=h=0$. A basis is given by the $L_t-1$ states
(we write them explicitly for $L_t=5$, i.e., $L=3$)
\begin{align}
  & |1\rangle = |{\rm k}_{3}\rangle = |1,1,1,1,-1\rangle, \nonumber \\
  & |2\rangle = |{\rm k}_{2}\rangle = |1,1,1,-1,-1\rangle, \nonumber \\
  & |3\rangle = |{\rm k}_{1}\rangle = |1,1,-1,-1,-1\rangle, \nonumber \\
  & |4\rangle = |{\rm k}_{0}\rangle = |1,-1,-1,-1,-1\rangle,  
  \label{basis-OBF}
\end{align}
where $|s_0,\ldots, s_{L+1}\rangle$ is an eigenstate of
$\hat{\sigma}_x^{(1)}$ with eigenvalue $s_x$, for all $x$'s.

The Hamiltonian restricted to one-kink states takes a tridiagonal
form.  The only nonvanishing elements are
\begin{eqnarray} 
   H_{nn} &=& E_0 + 2 h n - h L_t, \nonumber \\
   H_{nm} &=& g,  \qquad  n-m = \pm 1 \,,
   \label{Ham1K}
\end{eqnarray}
where $1 \le n,m \le L_t-1$, and $E_0$ is the energy of the states for
$g=h=0$.  Note that here we have changed the sign of $g$, with respect
to Eq.~\eqref{hedef}. As the model is invariant under $g\to-g$, the
results for the spectrum are independent of this choice.

The same analysis applies to the one-kink levels for a chain of length
$L_t = L$ with OBC.  In this case, for $g = h= 0$, the ground state is
doubly degenerate.  A basis is provided by the states that are fully
magnetized, i.e., $|+1,\ldots,+1\rangle$ and $|-1,\ldots,-1\rangle$,
in the same notation as above. For nonvanishing $g$ and/or $h$, the
degeneracy is lifted with a gap of order $g^L$ (for $h= 0$) or of
order $hL$ for small values of $h$. For $g=h=0$, the first excited
level consists of the $(2L-2)$ kink states. The degeneracy is
partially lifted by considering finite values of $g$ and/or $h$.
Since the Hamiltonian is invariant under space reflections, $\hat
\sigma_x^{(\alpha)} \to \hat \sigma_{L+1-x}^{(\alpha)}$ ($x =
1,\ldots,L$), we can divide the one-kink states in two sectors, with
basis (again we set $L_t=5$)
\begin{align}
  & |1\rangle_\pm = \tfrac{1}{\sqrt{2}}
  \left(|1,1,1,1,-1\rangle \pm |-1,1,1,1,1\rangle\right), \\
  & |2\rangle_\pm = \tfrac{1}{\sqrt{2}} \left(|1,1,1,-1,-1\rangle
  \pm |-1,-1,1,1,1\rangle\right),
  \nonumber \\
  & |3\rangle_\pm = \tfrac{1}{\sqrt{2}} \left(|1,1,-1,-1,-1\rangle
  \pm |-1,-1,-1,1,1\rangle\right),
  \nonumber \\
  & |4\rangle_\pm = \tfrac{1}{\sqrt{2}} \left(|1,-1,-1,-1,-1\rangle
  \pm |-1,-1,-1,-1,1\rangle\right). 
  \nonumber 
\end{align}
States $|i\rangle_\pm$ satisfy $\hat U |i\rangle_\pm = \pm
|i\rangle_\pm$ for all $i$, where $\hat U$ generates the space
reflections.  The Hamiltonian restricted to both sectors is again
given by the matrix reported in Eq.~\eqref{Ham1K}.

The spectrum of the restricted Hamiltonian~\eqref{Ham1K} can be
derived from the results reported in the Appendix of
Ref.~\cite{PRV-25}: one should simply replace $g$ with $g/2$ and $L$
with $L_t$.  The energies are given by
\begin{equation}
  {E}_n = E_0 - h L_t - 2 h \nu_n(z),
  \label{En1}
\end{equation}
where $\nu_n(z)$ satisfies the equation 
\begin{equation}
  J_{\nu}(z) = 0.
\end{equation}
Here $z = g/h$ and $J_{\nu}(z)$ is a Bessel function of the first
kind~\cite{GraRi}. Corrections to Eq.~(\ref{En1}) are exponentially small
in the size.  In the finite-size limit, in which $h\to 0$, $L \to
\infty$ at fixed $hL$, the lowest energy levels are given by (here we
can simply replace $L_t$ with $L$)
\begin{equation}
  E_n \approx E_0 - h L 
  - 2 g \left[ 1 - {|\alpha_n|\over 2}
    \left ({2 h \over g}\right)^{2/3}\right] ,
  \label{Enapprox}
\end{equation}
where $\alpha_n$ are the zeroes of the Airy function
$\mathop{\hbox{Ai}}(z)$.  The smallest zeroes correspond to $\alpha_n
= -2.33811,-4.08795,-5.52056, -6.78671, -7.94413$ for $n=1,2,3,4,5$.
Note that the nonanalytic term is of the order of $L^{-2/3}$ at fixed $h L$.
Corrections are of order $1/L$.

It is important to stress that the asymptotic result (\ref{Enapprox})
only holds in the limit $h \to 0$, $L\to \infty$ for fixed (and not
too large, as we explain below) values of $hL$. More precisely, it
does not hold for $h \to 0$ at fixed $L$, since for finite sizes the
behavior is analytic in $h$. In this limit, the magnetic-field
corrections are of order $h^2$ at fixed $L$, because of the symmetry
under $h\to -h$.  This type of behavior should be observed when the
magnetic energy, of order $h L$, is much smaller than the splitting of
the levels due to the transverse field, that is of the order of
$g/L^2$, i.e., for $h \ll g L^{-3}$. Equation~(\ref{Enapprox}) instead
applies for $h\gg g L^{-3}$. Indeed, if this condition holds, the
magnetic energy $hL$ is much larger than the correction term of order
$g (h/g)^{2/3}$.  Finally, let us note that, since we use perturbation
theory, the magnetic energy $hL$ should be small compared with the
spacing of the levels of the full theory for $g=h=0$. This requires $h
L \ll 4 J$.

The longitudinal magnetization $M_n$ associated with the $n^{\rm th}$
state follows immediately from Eq.~(\ref{Enapprox}).  Using the
Hellmann-Feynman theorem, we obtain
\begin{equation}
 M_n = - {1\over L} {\partial {E}_n\over \partial h}  
   = 1 - {4 |\alpha_n|\over 3} \left({g \over 2 hL}\right)^{1/3} L^{-2/3}.
\label{scalingM2}
\end{equation}
Again, we stress that the result (\ref{scalingM2}) does not hold for
$h\to0$ at fixed $L$.  In the latter case, $M_n\sim h$ for small
values of $h$. As a final remark, since the zeroes $\alpha_n$ of the
Airy function scale as $n^{2/3}$ for $n$ not too small, the effective
length scale that controls the corrections at fixed $hL$ is $L/n$,
implying that larger and larger lattice sizes are needed to observe
the asymptotic behavior of the energy or of the magnetization of the
$n{\rm th}$ kink level, as $n$ increases.

\begin{figure*}[!t]
  \includegraphics[width=0.45\textwidth]{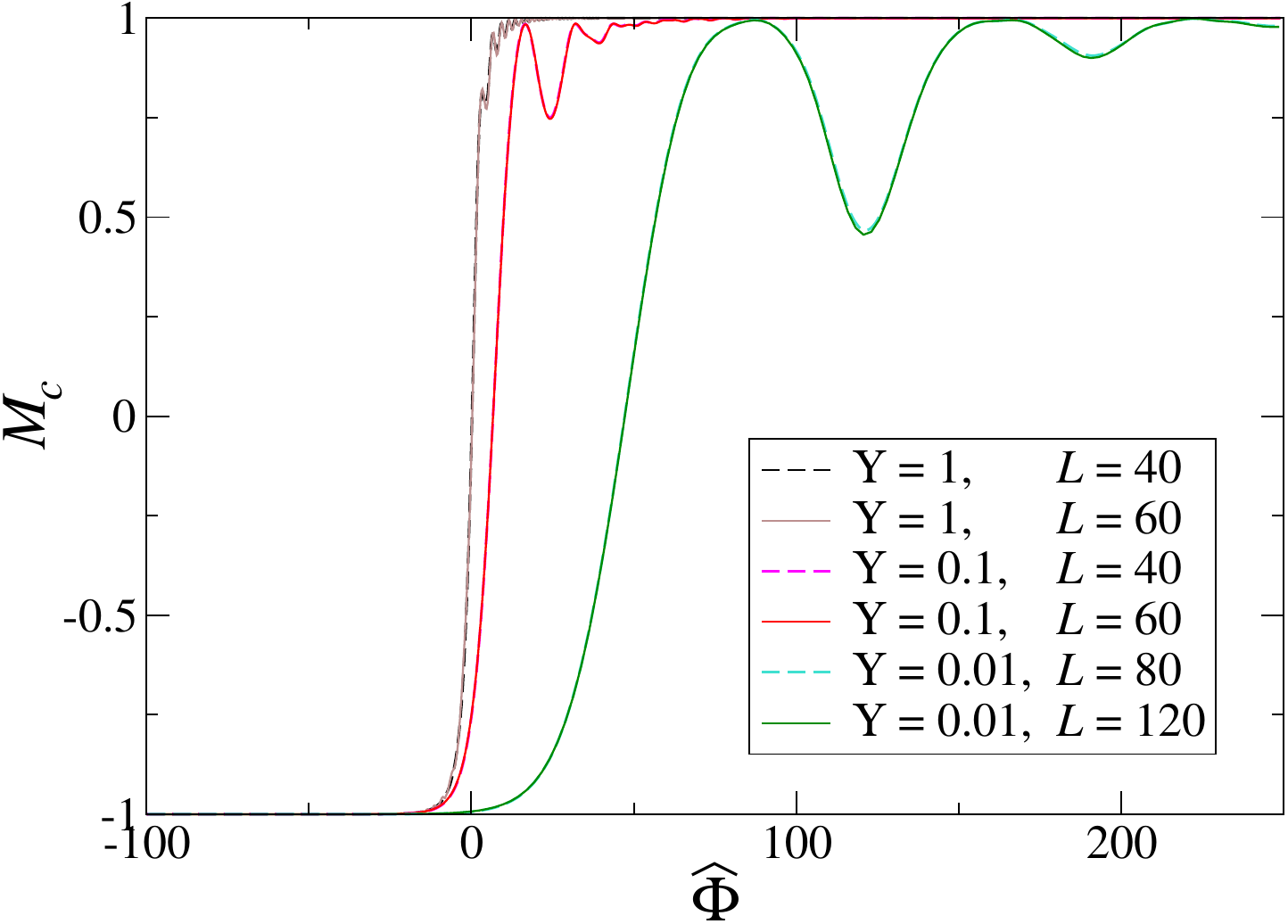} \hspace{4mm}
  \includegraphics[width=0.45\textwidth]{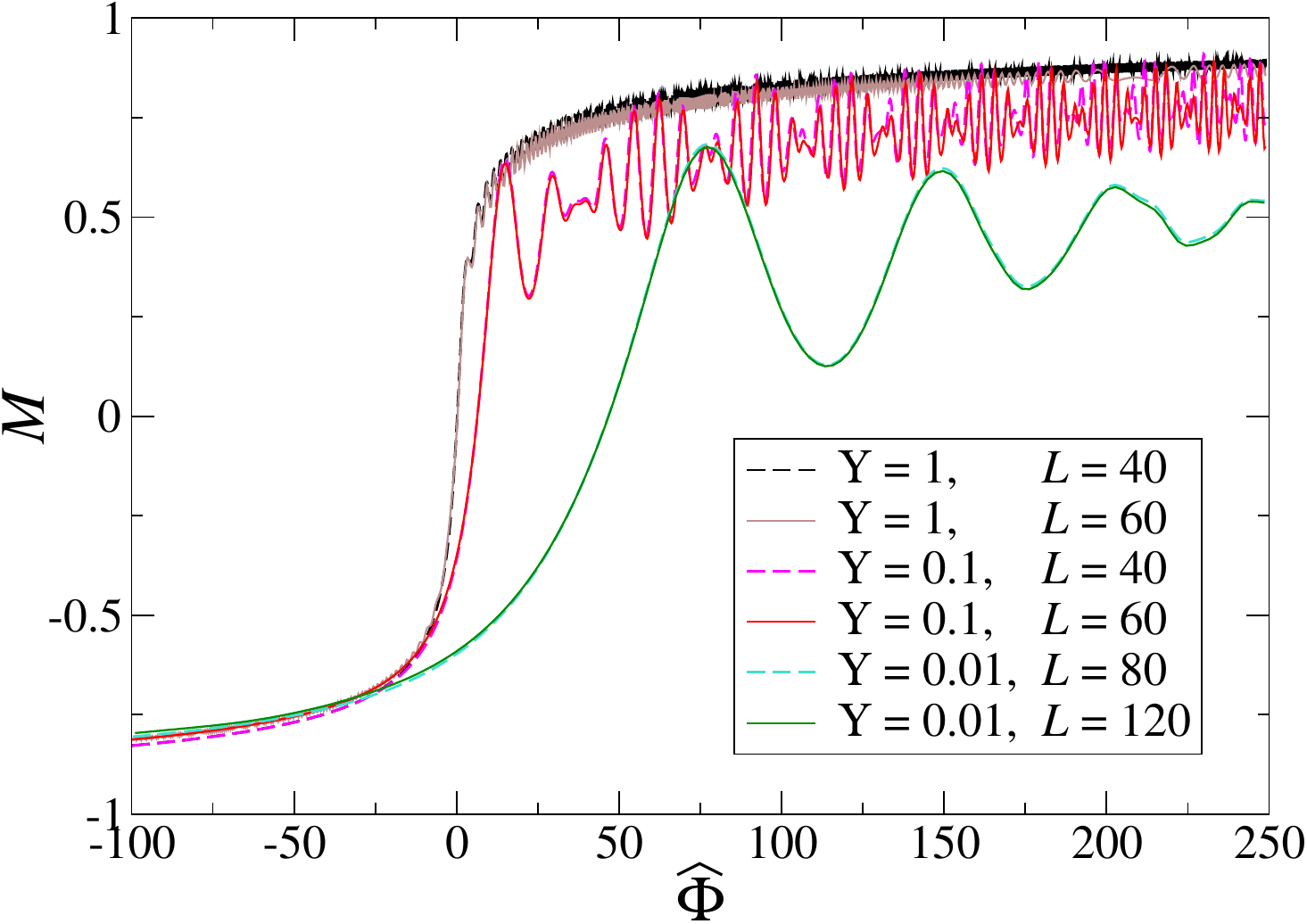} \vspace*{5mm} \\
  \includegraphics[width=0.46\textwidth]{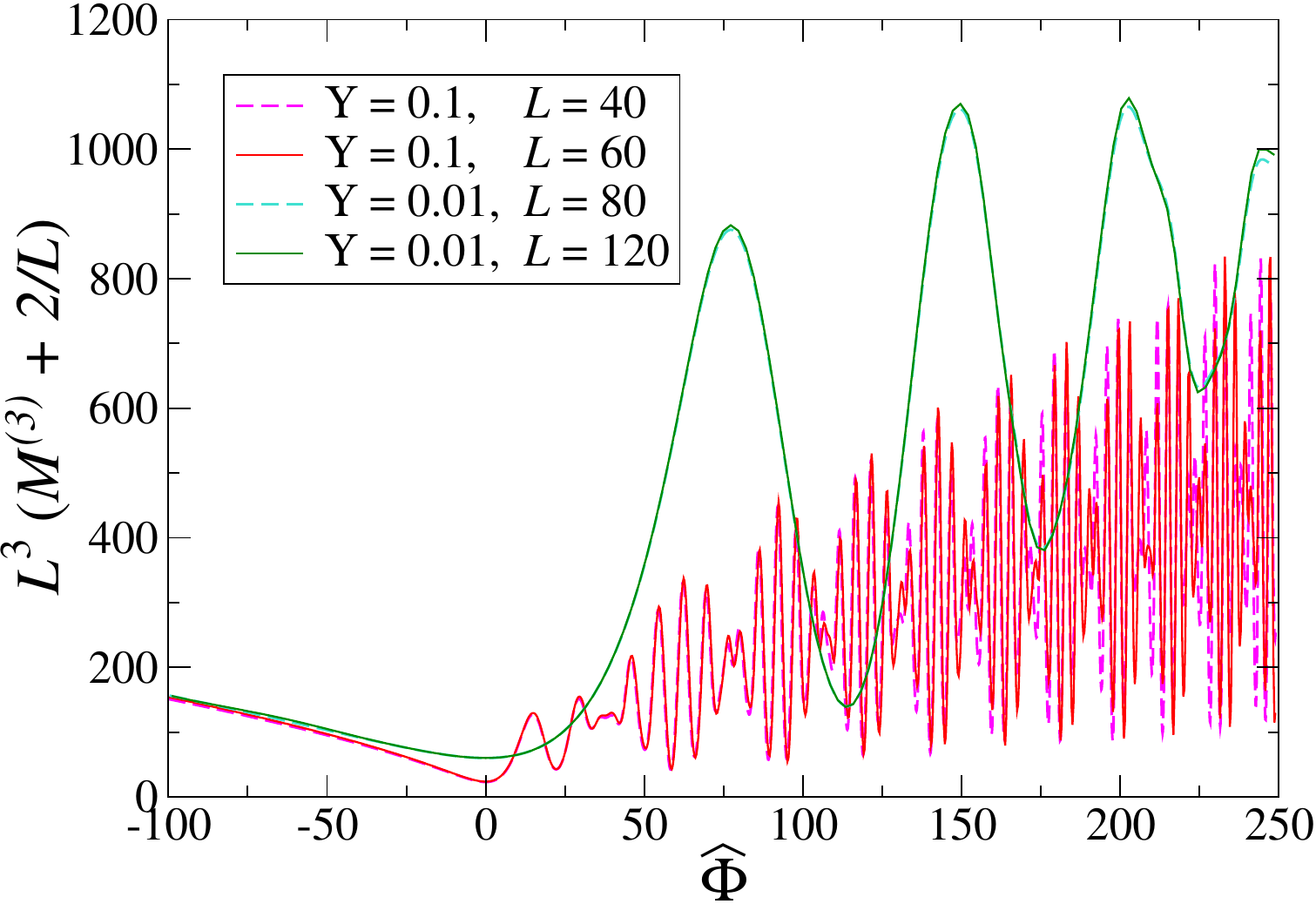} \hspace{3mm} 
  \includegraphics[width=0.46\textwidth]{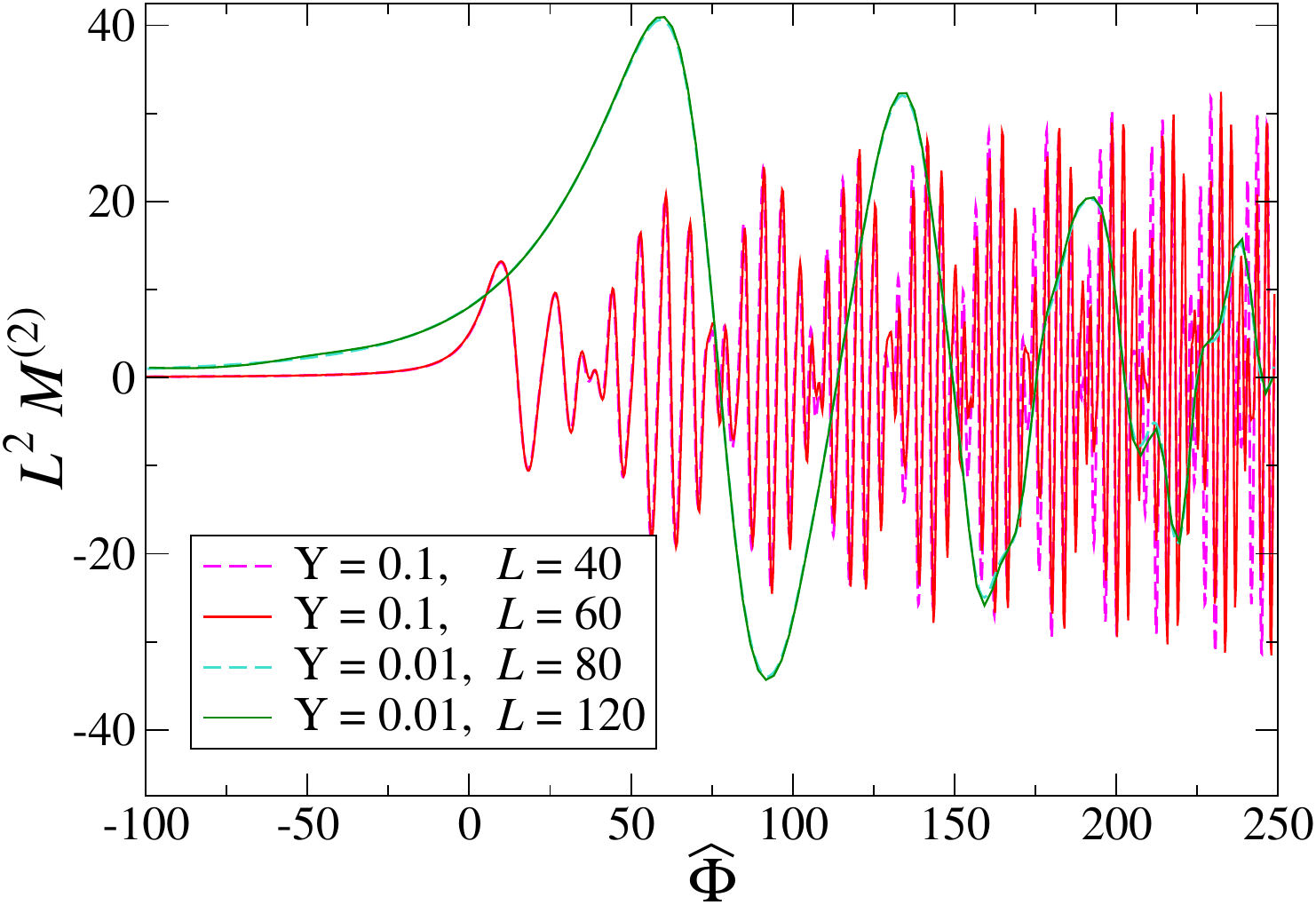}
  \caption{Scaling behavior of the magnetization as a function of
    $\widehat{\Phi}$, in the restricted one-kink subspace. Data are obtained
    for $g=0.05$ and different values of $\Upsilon$ and $L$.
    Top: results for the central $M_c$ (left) and the average $M$ (right)
    longitudinal component of the magnetization.  Bottom: rescaled
    components $M^{(3)}$ (left) and $M^{(2)}$ (right); these results
    provide evidence of the scaling relations~\eqref{scaling-Mtrasv}.}
  \label{scalingK}
\end{figure*}

\section{KZ Dynamics in the kink sector}
\label{AppOFSS}

We now focus on the OFSS limit for the KZ dynamics of chains with
OFBC.  This discussion also applies to OBC and PBC, provided that the
initial state of the dynamics is a one-kink or a two-kink state in the
two cases, respectively.

For OFBC and $g$ small, the gap is given by $\Delta(L) \approx 3 g
\pi^2/L^2$, so the relevant scaling variables, defined in
Eqs.~\eqref{katdef} and~\eqref{upsilondef}, are [note that, for $g\to
  0$, we have $m_0 = 1 + O(g^2)$]:
\begin{equation}
  \widehat{\Phi} = {2\over 3 \pi^2 g} h(t) L^3 \,, \qquad
  \Upsilon = {9 \pi^4 g^2\over 2} t_s L^{-5}.
\end{equation}
As $h \sim 1/L^3$ in the OFSS limit considered here---thus $hL\sim
1/L^2$---we can restrict the dynamics to the Hilbert space defined by
the basis introduced in Eq.~(\ref{basis-OBF}).

We have analyzed the components of the magnetization, defined in terms
of the local magnetization
\begin{equation}
  m_x^{(\alpha)}(t) = \langle\Psi(t)|
  \hat{\sigma}_{x}^{(\alpha)}|\Psi(t) \rangle ,
  \quad \alpha=1,2,3 ,
\end{equation}
where $x$ denotes a site of the chain.  In particular, we have
considered the components of the rescaled central and average
magnetization,
\begin{equation}
  M_c^{(\alpha)} \!=\! \frac{m_{L/2}^{(\alpha)} + m_{L/2+1}^{(\alpha)}}{2 m_0} ,
  \quad
  M^{(\alpha)} \!=\! \frac{1}{m_0 \, L} \sum_{x=1}^L m_x^{(\alpha)} ,
\end{equation}
so $M_c^{(1)}$ coincides with the rescaled central longitudinal
magnetization $M_c$ defined in Eq.~\eqref{MMcdef}, while $M_c^{(3)}$
is the corresponding magnetization component along the direction of
the transverse field $g$ (rescaled by a factor $m_0$, as well).  The
same applies to the average quantities $M^{(\alpha)}$.  We numerically
checked that the time dependence of the longitudinal magnetization
scales as discussed in Sec.~\ref{ofss}.  Moreover, we observe the
scalings
\begin{eqnarray}
 &&M^{(2)} \!=\! {1\over L^2} {\cal M}^{(2)} (\Upsilon,\widehat{\Phi}),
  \label{scaling-Mtrasv}\\
&&M^{(3)} \!=\! {a\over L} + {1\over L^3} {\cal M}^{(3)}
  (\Upsilon,\widehat{\Phi}),
  \nonumber
\end{eqnarray} 
where $a$ is a $t$- and $\Upsilon$-independent constant, which takes
the value $-2$.
These scaling Ans\"atze are confirmed by the numerical results
reported in Fig.~\ref{scalingK}. In all cases we observe an excellent
scaling, with results for different values of $L$ falling on top of
each other. We do not report results for $M^{(2)}$ and $M^{(3)}$ for
$\Upsilon = 1$, as the data show fast oscillations (with period in
$\widehat{\Phi}$ smaller than 0.1), which obscure the figure.

In general, we observe that scaling corrections increase as $\Upsilon$
is decreased. For instance, for $\Upsilon = 0.01$, we observe
significant deviations for $L = 20$ and a stable $L$-independent
behavior is only obtained, on the scale of the figure, for $L\gtrsim
50$. Second, the large-$\widehat{\Phi}$ behavior is always
characterized by oscillations that increase in amplitude and decrease
in frequency, as $\Upsilon$ decreases.  Finally, a transverse
magnetization develops as the system crosses the transition: both
$M^{(2)}$ and $M^{(3)}$ are nonvanishing for $\widehat{\Phi} > 0$.
This effect disappears as $L\to \infty$.

The observed scaling behavior can be predicted analytically by
considering the exact evolution equations for the components of the
magnetization, computed in the restricted model. Since, for any
quantity $\hat{A}$ and state $|\Psi(t)\rangle$ we have
\begin{equation}
  {{\rm d}\over {\rm d}t} \langle \Psi(t)|\hat{A}|\Psi(t)\rangle = 
  i \langle \Psi(t)|[\hat H,\hat{A}]|\Psi(t)\rangle,
\end{equation}
we obtain 
\begin{eqnarray}
  {{\rm d} M^{(1)}(t)\over {\rm d}t} &=& 2 g \, M^{(2)}(t), \label{eqMz} \\ 
  {{\rm d} M^{(3)}(t)\over {\rm d}t} &=& 2 h(t) \, M^{(2)}(t), \label{eqMx} \\
  {{\rm d} M^{(2)}(t)\over {\rm d}t} &=& -2 h(t) \, M^{(3)}(t) + 2 g \, M_b(t),
  \label{eqMy}
\end{eqnarray}
where
\begin{equation}
  M_b(t) = \langle \Psi(t) | \hat{M}_b | \Psi(t)\rangle,
  \;\; \hat{M}_b = \hbox{diag}(1,0,\ldots,0,-1),
  \label{mbdef}
\end{equation}
written in the basis reported in Eq.~(\ref{basis-OBF}). This set of
equations does not close, because of the presence of the new operator
$\hat{M}_b$. Nevertheless, it allows us to explain some general
properties of the scaling behavior.  By replacing $t$ with
$\widehat{\Phi}$, Eq.~(\ref{eqMz}) can be rewritten as
\begin{equation}
  {{\rm d}M^{(1)}(t)\over {\rm d}\widehat{\Phi}} =
  {2\over 3\pi^2} \Upsilon L^2 M^{(2)}(t),
  \label{eqMx_FSS}
\end{equation}
which is consistent with the general scaling behavior of $M^{(1)}$
reported in Eq.~(\ref{mtsl}) only if $M^{(2)}$ scales as $1/L^2$ in
the OFSS limit, i.e., it satisfies the scaling relation
(\ref{scaling-Mtrasv}).  Analogously, Eq.~(\ref{eqMx}) implies
\begin{equation}
  {{\rm d}M^{(3)}(t)\over {\rm d}\widehat{\Phi}}
  = \widehat{\Phi} \Upsilon {1\over L^3} [L^2 M^{(2)}(t)] ,
\end{equation}
which implies that the derivative of $M^{(3)}(t)$ should scale as
$1/L^3$, consistently with Eq.~(\ref{scaling-Mtrasv}).  Finally,
Eq.~(\ref{eqMy}) can be rewritten as
\begin{equation}
  {{\rm d}[L^2 M^{(2)}(t)]\over {\rm d}\widehat{\Phi}}
  = -\Upsilon \left[\widehat{\Phi} L M^{(3)}(t)
    - {2\over 3\pi^2}  L^4 M_b(t)\right],
  \label{eqMy_FSS}
\end{equation}
which implies 
\begin{equation}
  M_b(t) = {1\over L^4} {\cal M}_b(\widehat{\Phi},\Upsilon).
\end{equation} 
Moreover, $M^{(3)}$ should scale as $1/L$.  Since the derivative of
$M^{(3)}$ scales as $1/L^3$, the term of order $1/L$ should be
independent of $\widehat{\Phi}$, i.e., $M^{(3)}(t) = a/L + O(1/L^3)$,
where $a$ does not depend on $\widehat{\Phi}$, i.e., on $t$.  To
estimate $a$ we consider the limiting behavior for $\widehat{\Phi}\to
-\infty$, which can also be obtained by considering first the limit
$L\to \infty$, $t_s\to \infty$, $t\to -\infty$ at fixed $\Upsilon$ and
fixed negative $h_0 = t/t_s$ and then taking the limit $h_0\to
0$. However, for fixed $h_0< 0$ the system is gapped, so for
$t_s\to\infty$ and any large value of $L$ the state $|\Psi(t)\rangle$
coincides with the ground state of the system.  It follows that the
constant $a$ can be computed by considering the ground-state
magnetization in the $z$-direction for $h = 0^-$ in the large-$L$
limit. Using the Hellmann-Feynman theorem, we should evaluate
$\partial E_0/\partial g$ for $h \to 0^-$, where $E_0$ is the
ground-state energy.  Since $E_0 = - 2 g + O(1/L^2)$ in the limit
$h\to 0$, we obtain $M^{(3)} = - 2/L$ for large values of $L$. It
follows $a = -2$.  In the OFSS limit Eq.~(\ref{eqMy_FSS}) can thus be
rewritten as
\begin{equation}
  {{\rm d}{\cal M}^{(2)}\over {\rm d}\widehat{\Phi}}
  = -\Upsilon \left(a \widehat{\Phi} - {2\over 3\pi^2}  {\cal M}_b\right),
  \label{eqMy_FSS_2}
\end{equation}
It is interesting to compute the behavior of the OFSS scaling
functions in the limit $\Upsilon \to 0$. As discussed in
Sec.~\ref{abcsec}, scaling functions should only depend on the
variable $W = \widehat{\Phi} \Upsilon^{3/5}$.  In this limit
Eq.~(\ref{eqMy_FSS_2}) can be rewritten as
\begin{equation}
  {{\rm d}{\cal M}^{(2)}\over {\rm d}W} = - a W \Upsilon^{-1/5}
  + {2 \over 3 \pi^2} {\cal M}_b \Upsilon^{2/5}.
\end{equation}
A finite limit is thus obtained if
\begin{equation}
{\cal M}^{(2)} \approx
\Upsilon^{-1/5} \widehat{\cal M}^{(2)}(W),
\;\;
{\cal M}_b \approx
\Upsilon^{-3/5} \widehat{\cal M}_b(W).
\label{condsca}
\end{equation}
Equation~(\ref{eqMx_FSS}) allows
us to derive the behavior of the derivative of the scaling function
${\cal M}^{(1)}(\widehat{\Phi},\Upsilon)$. Indeed, it can be written
as
\begin{equation}
  {{\rm d}{\cal M}^{(1)}\over {\rm d}W} = {2\over 3 \pi^2}
  \Upsilon^{2/5} {\cal M}^{(2)} = {2\over 3 \pi^2} \Upsilon^{1/5}
  \widehat{\cal M}^{(2)}(W),
\end{equation}
where the last equality holds for $\Upsilon \to 0$. This implies that
the derivative of ${\cal M}^{(1)}$ vanishes as $\Upsilon^{1/5}$ for
$\Upsilon \to 0$.  In turn, this implies
\begin{equation}
  {\cal M}^{(1)}(\Upsilon,\widehat{\Phi}) = -1 + \Upsilon^{1/5}
  \widehat{\cal M}^{(1)}(W),
  \label{smallUpsilon}
\end{equation}
where we have used $M^{(1)}(\Upsilon,\widehat{\Phi} = 0) = - 1$ in the
OFSS limit. Equation~(\ref{smallUpsilon}) should hold for any $g < 1$.

\begin{figure*}[t]
  \includegraphics[width=8truecm,angle=0]{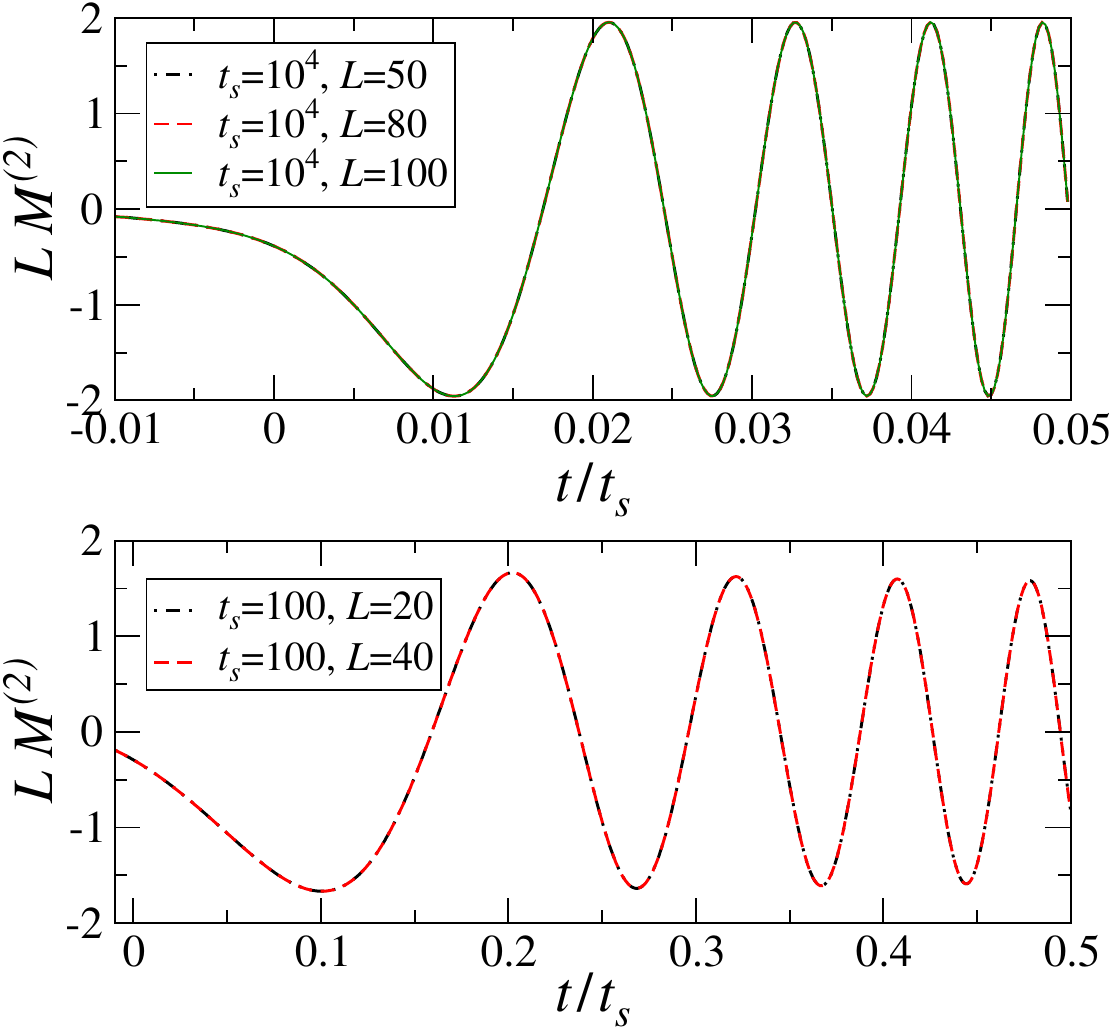}
  \includegraphics[width=8truecm,angle=0]{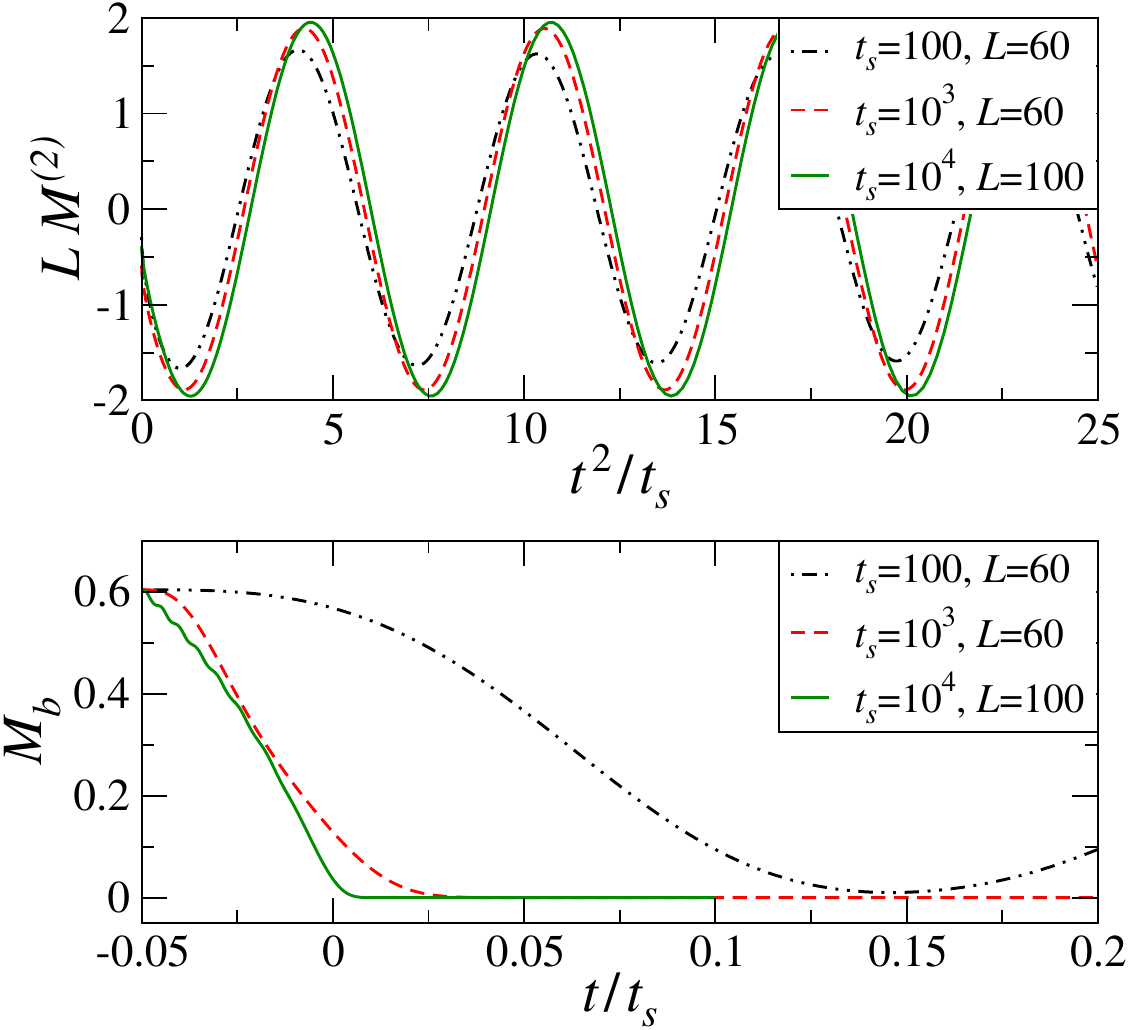}
  \caption{Scaling behavior of $M^{(2)}$ and $M_b$, in the
    infinite-volume limit.  In the two panels on the left, we report
    $L M^{(2)}$ vs $t/t_s$, for two different values of $t_s$ and
    several values of $L$. Curves for different values of $L$ cannot
    be distinguished, as data fall on top of each other on the scale of
    the figure. In the upper panel on the right, we report $L M^{(2)}$
    vs $t^2/t_s$ for three different values of $t_s$.  In the lower
    panel on the right, 
    we report $M_b$ vs $t/t_s$ for three different values of
    $t_s$. For each $t_s$, we only report results for one value of $L$,
    as data are $L$-independent on the scale of the figure.}
  \label{scal-fixedts}
\end{figure*}

The previous scaling relations hold in the OFSS limit, in which $t_s$
is of the order of the typical time scale $T(L)$ defined in
Eq.~(\ref{upsilondef}).  We now wish to discuss the behavior for $t_s
\ll T(L)$, which is strongly nonadiabatic. For $h < 0$, the ground
state is essentially given by the state (with a small $g$ correction)
\begin{equation}
  | {\rm k}_0\rangle = |1,-1,...,-1,-1\rangle,
  \label{kldef}
\end{equation}
where the first and last single spin states refer to the fixed
boundaries at $x=0$ and $x=L+1$ respectively, thus $|{\rm k}_0\rangle$
corresponding to the state $|L_t-1\rangle$ defined in
Eq.~(\ref{basis-OBF}).  The corresponding longitudinal magnetization
is given by $M^{(1)} \approx -1$.  As the system crosses the
transition $h=0$, this state becomes the highest-energy kink state and
the system starts to make transitions to reach lower-energy states.
However, if $t_s$ is too small, the system makes only a small finite
number of transitions, ending in a new state with an $x$-component of
the magnetization $M^{(1)} \approx -1 + O(1/L)$. This result follows
from the numerical observation that, if we take the limit $L\to
\infty$ at fixed $t_s$, the magnetization $M^{(2)}$ satisfies the
scaling behavior $M^{(2)}(t,t_s,L) = \widetilde{\cal
  M}^{(2)}(t/t_s,t_s)/L$. This is evident from Fig.~\ref{scal-fixedts} 
(left panels), where we report $L M^{(2)}$ for two different values
of $t_s$ and several values of $L$: no $L$-dependence is visible on the
scale of the figure.  By integrating Eq.~(\ref{eqMz}) we obtain
\begin{equation}
  M^{(1)}(t,t_s,L) = 
  -1 + {2g\over L} \int_0^t dt\, \widetilde{\cal M}^{(2)}(t/t_s,t_s),
\end{equation}
where we assume $M^{(1)}(0,t_s,L) = - 1 + O(1/L^2) \approx - 1$ for
$L\to\infty$.  Albeit $M^{(1)}\to -1$ in the limit $L\to \infty$ for
all values of $t$ and $t_s$, the system ends in a superposition of
states that significantly depend on the value of $t_s$.  This is
signaled by the behavior of $M_b(t,t_s,L)$, which provides a
quantitative estimate of $|\langle {\rm k}_0|\Psi(t)\rangle|^2$ (note
that $|\langle {\rm k}_L |\Psi(t)\rangle|^2$, where $|{\rm k}_L\rangle
= |1,1,...,1,-1\rangle$, always vanishes).  This quantity is
essentially independent of $L$ in the regime $t_s \ll T(L)$ but has a
significant dependence on $t_s$ as shown in Fig.~\ref{scal-fixedts}
(lower panel on the right), where we report $M_b$ for different values
of $t_s$. While for $t_2 = 10^2$, $M_b$ is nonzero, for $t_s = 10^4$,
$|M_b|\lesssim 10^{-7}$ for $t/t_s \gtrsim 0.015$, so $|{\rm
  k}_0\rangle$ [this state is defined in Eq.~(\ref{kldef})] is no
longer relevant for the state $|\Psi(t)\rangle $ of the system. The
same occurs for $t_s = 10^3$ and is a general feature of the dynamics
for $t_s \gtrsim 10^3$ in the infinite-volume limit. For these values
of $t_s$, we can approximate $M_b\approx 0$ when $t>t^*$, where $t^*$
is a $t_s$-dependent cutoff time. Under these conditions, we can
integrate Eqs.~\eqref{eqMx} and~\eqref{eqMy}, obtaining
\begin{equation}
  M^{(2)}(t,t_s,L) = {a\over L} \cos\left({t^2\over t_s} + \varphi\right),
  \label{My-infvol}
\end{equation}
where $a$ and $\varphi$ may in principle depend on $t_s$. However, as
shown in Fig.~\ref{scal-fixedts} (upper panel on the right), results appear
to be independent of $t_s$ when plotted vs $t^2/t_s$, indicating
that both $a$ and $\varphi$ are independent of $t_s$ (more precisely,
$a$ and $\varphi$ have a finite limit as $t_s\to\infty$).
Eq.~(\ref{My-infvol}) implies the existence of a scaling limit in
terms of $\hat{\tau}=t/\sqrt{t_s}$.  Again, we can use the scaling
results for $M^{(2)}$ to infer the scaling behavior of $M^{(1)}$:
\begin{equation}
  M^{(1)}(t,t_s,L) \!=\! M^{(1)}(t^*,t_s,L) + {a \sqrt{t_s}\over L}
  \!\!\!  \int_{\hat{\tau}^*}^{\hat{\tau}} d\tau \cos(\tau^2+\phi),
\end{equation}
where $\hat{\tau}^*=t^*/\sqrt{t_s}$. These results explain the rapid
oscillations of the components of the magnetization with an effective
frequency that increases with increasing $t/t_s$ and with an amplitude
that increases with $t_s$ (as $t_s^{1/2}$) and decreases with $L$ (as
$1/L$).

It is important to stress that the behavior discussed in this appendix
only applies for a small range of values of $h$, i.e., for $h\ll 1/L$.
Therefore, this discussion is not relevant for the infinite-volume
behavior in terms of the variable $\Omega(t)$, presented in
Sec.~\ref{ofbcsec}, which occurs for values of $h(t)$ at which the
one-kink approximation is no longer valid.

\end{document}